\documentclass[aps,twocolumn,superscriptaddress,amsmath,longbibliography,nofootinbib]{revtex4-1}
\usepackage[pdftex]{graphicx,color}
\usepackage{soul}
\usepackage{amsfonts}
\usepackage{amssymb}
\usepackage{amsmath}
\usepackage{amsthm}
\usepackage{bm}
\usepackage{braket}
\usepackage{enumitem}

\begin{document}

\title{Disordered Lieb-Robinson bounds in one dimension}

\author{Christopher L. Baldwin}
\affiliation{Joint Quantum Institute, University of Maryland, College Park, MD 20742, USA}

\author{Adam Ehrenberg}
\affiliation{Joint Quantum Institute, University of Maryland, College Park, MD 20742, USA}
\affiliation{Joint Center for Quantum Information and Computer Science, NIST/University of Maryland, College Park, MD 20742, USA}

\author{Andrew Y. Guo}
\affiliation{Joint Quantum Institute, University of Maryland, College Park, MD 20742, USA}
\affiliation{Joint Center for Quantum Information and Computer Science,
NIST/University of Maryland, College Park, MD 20742, USA}

\author{Alexey V. Gorshkov}
\affiliation{Joint Quantum Institute, University of Maryland, College Park, MD 20742, USA}
\affiliation{Joint Center for Quantum Information and Computer Science, NIST/University of Maryland, College Park, MD 20742, USA}
\affiliation{National Institute of Standards and Technology, Gaithersburg, MD 20899, USA}

\date{\today}

\begin{abstract}

By tightening the conventional Lieb-Robinson bounds to better handle systems which lack translation invariance, we determine the extent to which ``weak links'' suppress operator growth in disordered one-dimensional spin chains.
In particular, we prove that ballistic growth is impossible when the distribution of coupling strengths $\mu(J)$ has a sufficiently heavy tail at small $J$, and identify the correct dynamical exponent to use instead.
Furthermore, through a detailed analysis of the special case in which the couplings are genuinely random and independent, we find that the standard formulation of Lieb-Robinson bounds is insufficient to capture the complexity of the dynamics --- we must distinguish between bounds which hold for \textit{all} sites of the chain and bounds which hold for a \textit{subsequence} of sites, and we show by explicit example that these two can have dramatically different behaviors.
All the same, our result for the dynamical exponent is tight, in that we prove by counterexample that there cannot exist any Lieb-Robinson bound with a smaller exponent.
We close by discussing the implications of our results, both major and minor, for numerous applications ranging from quench dynamics to the structure of ground states.

\end{abstract}

\maketitle

\section{Introduction} \label{sec:introduction}

Lieb-Robinson (LR) bounds, named after Ref.~\cite{Lieb1972Finite}, have proven to be a valuable mathematical tool for many-body physics and quantum information theory~\cite{Hastings2004LiebSchultzMattis,Nachtergaele2006LiebRobinson,Hastings2006Spectral,Bravyi2006LiebRobinson,Osborne2006Efficient,Schuch2011Information,Abanin2015Exponentially,Shiraishi2017Efficiency,Eldredge2017Fast,Deshpande2018Dynamical,Haah2018Quantum,Bentsen2019Fast,Nachtergaele2019QuasiLocality}.
Conceptually, they provide hard constraints on the extent to which correlations of any type can spread through a many-body lattice system (broadly termed ``operator spreading'').
Numerous applications, from rigorous results on many-body ground states to lower bounds on the runtime of quantum protocols, can be found in the above references.

Given their utility, LR bounds have been both generalized and specialized in multiple ways: leveraging the commutativity and graph structure of interactions~\cite{PremontSchwarz2010aLiebRobinson,Wang2020Tightening,Chen2021Operator}, allowing for long-range interactions~\cite{FossFeig2015Nearly,Tran2019Locality,Chen2019Finite,Kuwahara2020Strictly,Tran2021LiebRobinson}, and considering open systems~\cite{Poulin2010LiebRobinson,Descamps2013Asymptotically,Sweke2019LiebRobinson,Guo2021Clustering}, to name a few.
However, one ingredient that has been noticeably absent is disorder, or more generally a lack of translation invariance.
To be fair, the conventional bounds do allow for non-translation-invariant interactions, but there have been no studies assessing the tightness of the resulting bounds (and we shall find that they are far from tight).
Certain works have considered related topics --- Refs.~\cite{Burrell2009Information,Gebert2022LiebRobinson} study the effects of disordered local terms (albeit with uniform interactions), and Ref.~\cite{Chen2021Concentration} derives bounds for random Hamiltonians with statistical translation invariance\footnote{\label{note:typicality}Ref.~\cite{Chen2021Concentration} does in fact allow for non-translation-invariant interactions, but in the same manner as do the conventional bounds. It would be worthwhile to investigate whether the results of Ref.~\cite{Chen2021Concentration} can be tightened much as we do here.}.
Others have calculated bounds for \textit{specific} (often free-fermion-integrable) systems~\cite{Burrell2007Bounds,Hamza2012Dynamical,Gebert2016On,Elgart2018Manifestations}.
While interesting on their own merits, none of these quite address the question with which we concern ourselves here --- to what extent is operator spreading (as constrained by LR bounds) necessarily suppressed by non-translation-invariant interactions?

Non-translation-invariant systems are known to exhibit a variety of phenomena not found in their translation-invariant counterparts.
Examples include spin glass phases (both static and dynamic)~\cite{Fischer1991,Mezard2009}, localization~\cite{Lee1985Disordered,Kramer1993Localization}, and Griffiths effects~\cite{Vojta2010Quantum}.
Disordered fermionic models have been used to help understand quantum dots and strongly correlated metals as well~\cite{Song2017Strongly,Altland2019Quantum,Altland2019Sachdev}.
Of particular (and somewhat controversial) interest is the phenomenon of many-body localization (MBL)~\cite{Basko2006Metal,Huse2014Phenomenology,Nandkishore2015Many,Abanin2019Colloquium}, whose existence remains under debate~\cite{Imbrie2016On,Suntajs2020Quantum,Abanin2021Distinguishing,Sels2021Dynamical,Crowley2020Constructive}.
Given the challenges inherent in studying not only MBL but disordered quantum many-body systems in general, it is all the more important to identify rigorous constraints such as those which LR bounds supply (although to be clear, our results and LR bounds in general are not strong enough to resolve the questions surrounding MBL, as we explain in Sec.~\ref{sec:summary}).

In the present work, we initiate the study of non-translation-invariant LR bounds by considering arguably the simplest (but still quite rich) situation: one-dimensional chains with nearest-neighbor interactions.
An essential feature of such systems is the importance of ``weak links'', i.e., atypically weak interactions.
We develop the machinery for analyzing these systems, prove that the LR bounds thus obtained are in a certain sense optimal, and use the results to place constraints on various physical properties and processes.

Sec.~\ref{sec:summary} summarizes our results in conceptual terms.
We have aimed to make it sufficiently self-contained so that readers primarily interested in our conclusions and willing to forgo the derivations should be able to read Sec.~\ref{sec:summary} on its own.
Sec.~\ref{sec:definitions} then gives precise definitions of all quantities involved in our analysis.
Sec.~\ref{sec:general_bound} derives LR bounds for general non-translation-invariant systems, and Sec.~\ref{sec:disordered_bound} specializes to the case of random couplings.
Sec.~\ref{sec:applications} lastly discusses some implications of our results, in particular related to: quench dynamics, topological order, heating rates, ground state correlations, and machine learning of local observables.

\section{Summary of results} \label{sec:summary}

\begin{figure}[t]
\centering
\includegraphics[width=1.0\columnwidth]{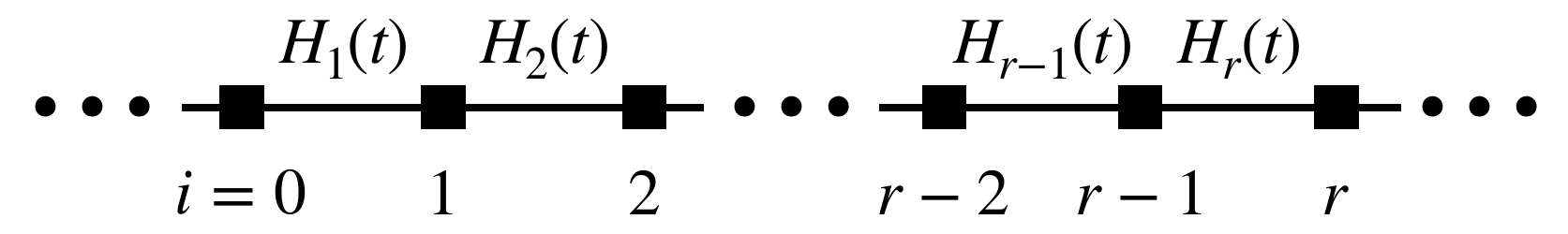}
\caption{Geometry of the systems studied in the present work, i.e., 1D chains with nearest-neighbor interactions. Sites are indicated by black squares, labelled by $i$. Terms of the Hamiltonian are indicated by solid lines, with $H_l(t)$ acting on the two sites connected by that link. Given a set $\{J_l\}$, we require that $\lVert H_l(t) \rVert \leq J_l$.}
\label{fig:lattice_layout}
\end{figure}

\begin{figure}
\centering
\includegraphics[width=1.0\columnwidth]{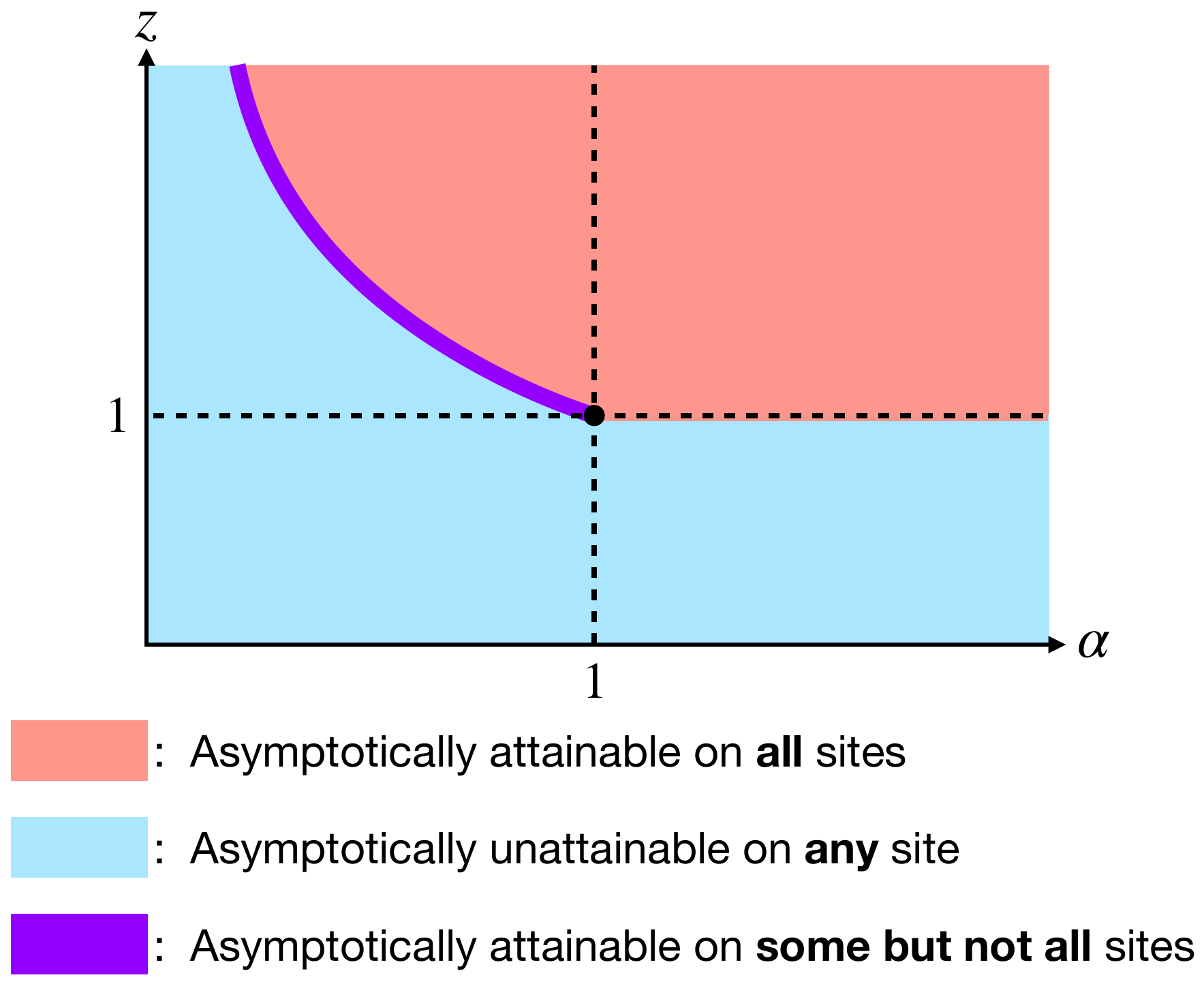}
\caption{The possibility of a Hamiltonian causing operator spreading at dynamical exponent $z$, i.e., in a time growing asymptotically as $t(r) \propto r^z$ to reach a large distance $r$. Horizontal axis is the exponent $\alpha$ characterizing the number of weak links in the couplings $\{J_l\}$ --- the fraction of links between sites 0 and $r$ having $J_l \leq J$ is assumed to go as $J^{\alpha}$ at small $J$ and large $r$. Vertical axis is $z$. Blue region is where no Hamiltonian with such $\alpha$ can reach \textit{any} large-distance site at such $z$. Red region is where a Hamiltonian exists with such $\alpha$ that reaches \textit{every} large-distance site at (or faster than) such $z$. The boundary between the two is given by $z_c(\alpha) = \max[1/\alpha, 1]$. In the special case of independent random couplings, the $\alpha > 1$ portion of the boundary is included in the red region, while the $\alpha < 1$ portion (shown in purple) is where a Hamiltonian exists that reaches \textit{a subsequence} of sites faster than $z$ but where no Hamiltonian can reach \textit{every} site at such $z$. Lastly, at point $(\alpha, z) = (1, 1)$, no Hamiltonian can reach every site ballistically but it is unknown whether any Hamiltonian can reach some sites ballistically.}
\label{fig:main_result}
\end{figure}

We determine the extent to which operator spreading in 1D nearest-neighbor chains is suppressed by weak links.
To be precise, we consider arbitrary $d$-state degrees of freedom (``qudits'') interacting via any Hamiltonian of the form $H(t) = \sum_l H_l(t)$, where the sum is over links of the chain and $H_l(t)$ acts on the two sites connected by link $l$ (although our analysis in fact allows for arbitrary local terms as well\footnote{\label{note:local_terms}Our results actually apply to any Hamiltonian of the form $H(t) = \sum_{l \in \Lambda} H_l(t) + \sum_{i \in \Omega} h_i(t)$, for \textit{arbitrary} local terms $h_i(t)$. The local terms can trivially be accounted for by first passing to the interaction picture with respect to them --- the terms $H_l(t)$ in this new frame have both the same support and the same norm as originally, hence our analysis applies equally well using them. That said, the behavior of any \textit{specific} system clearly can depend strongly on the local terms.}).
See Fig.~\ref{fig:lattice_layout}.
Unlike previous works, we assume that $\lVert H_l(t) \rVert \leq J_l$ where $J_l$ \textit{varies} from link to link ($\lVert \, \cdot \, \rVert$ denotes the operator norm throughout).
The ``weak'' links are those on which $J_l$ is much smaller than the typical value.

For arbitrary local operators $A_0$ and $B_r$ supported on sites 0 and $r$ respectively, and with $A_0^t$ denoting the evolution of $A_0$ over time $t$, our goal is to bound the quantity $\lVert [A_0^t, B_r] \rVert$ as tightly as possible, making use of the weak links in the set $\{J_l\}$.
We particularly focus on the asymptotic behavior at large $r$ and $t$.
The commutator $[A_0^t, B_r]$ is a standard and useful measure of operator spreading (as we describe in Sec.~\ref{subsec:definition_basis_strings}), but it is by no means the only such measure.
In fact, most of our calculations will involve a different quantity from which bounds on the commutator easily follow.

LR bounds are best viewed as applying to \textit{families} of Hamiltonians.
For each set of couplings $\{J_l\}$, we use $\mathcal{H}_J$ to denote the set of all Hamiltonians as described above consistent with $\{J_l\}$ (i.e., $\lVert H_l(t) \rVert \leq J_l$ for all links $l$ at all times $t$).
Our bounds are uniform among $H \in \mathcal{H}_J$, in the sense that they make no reference to any property of the Hamiltonian beyond the couplings $\{J_l\}$.
Thus while the results are extremely general, they may not be very tight for one specific system.
The ``tightness'' of LR bounds referred to in this paper is instead the existence of \textit{some} $H \in \mathcal{H}_J$ which saturates the bound.

As a consequence of their generality, LR bounds have little to say regarding, e.g., the existence of MBL --- the complete lack of operator spreading in strongly disordered time-independent systems~\cite{Basko2006Metal,Huse2014Phenomenology,Nandkishore2015Many,Abanin2019Colloquium}.
The same applies to other phenomena involving slow dynamics under a fixed Hamiltonian, such as activated processes in spin glasses~\cite{Chamon2002Separation,Bapst2013Quantum,Baldwin2018Quantum}.
Those $H \in \mathcal{H}_J$ that saturate our bounds will instead tend to be highly time-\textit{dependent} Hamiltonians, specifically designed to transmit information and better viewed as quantum circuits.
These are the types of systems for which LR bounds give a reasonably full picture.

Our first result, which holds for any possible set of couplings $\{J_l\}$, is an improvement on the conventional LR bound.
Whereas the standard analysis\footnote{\label{note:velocities}We are certainly not the first to improve on Eq.~\eqref{eq:conventional_LR_bound} --- see, e.g., Refs.~\cite{Wang2020Tightening,Chen2021Operator} for modifications that yield smaller LR velocities, sometimes even parametrically so. We have not attempted to marry our approach with theirs, but we expect that it can be done and leave this for future work.} leads to the result
\begin{equation} \label{eq:conventional_LR_bound}
\Big\lVert \big[ A_0^t, B_r \big] \Big\rVert \leq C \big\lVert A_0 \big\rVert \big\lVert B_r \big\rVert \left( \prod_{l=1}^r 4J_l \right) \frac{t^r}{r!},
\end{equation}
we show that one further has
\begin{equation} \label{eq:improved_LR_bound}
\Big\lVert \big[ A_0^t, B_r \big] \Big\rVert \leq C \big\lVert A_0 \big\rVert \big\lVert B_r \big\rVert \min_{\lambda} \left[ \left( \prod_{l \in \lambda} 4J_l \right) \frac{t^{|\lambda|}}{|\lambda|!} \right],
\end{equation}
where the minimization is over all subsets $\lambda$ of the links between sites 0 and $r$ ($|\lambda|$ denotes the size of the subset).
Even though much of what follows will be concerned with the asymptotic behavior, Eq.~\eqref{eq:improved_LR_bound} holds for all $r$ and $t$.
In these equations and throughout the entire paper, we use $C$ to denote any constant which does not depend on $r$ or $t$ and whose precise value is irrelevant to our conclusions.
Its value will often change between expressions (and primes/subscripts will differentiate such constants within the same expression).

\begin{figure}[t]
\centering
\includegraphics[width=1.0\columnwidth]{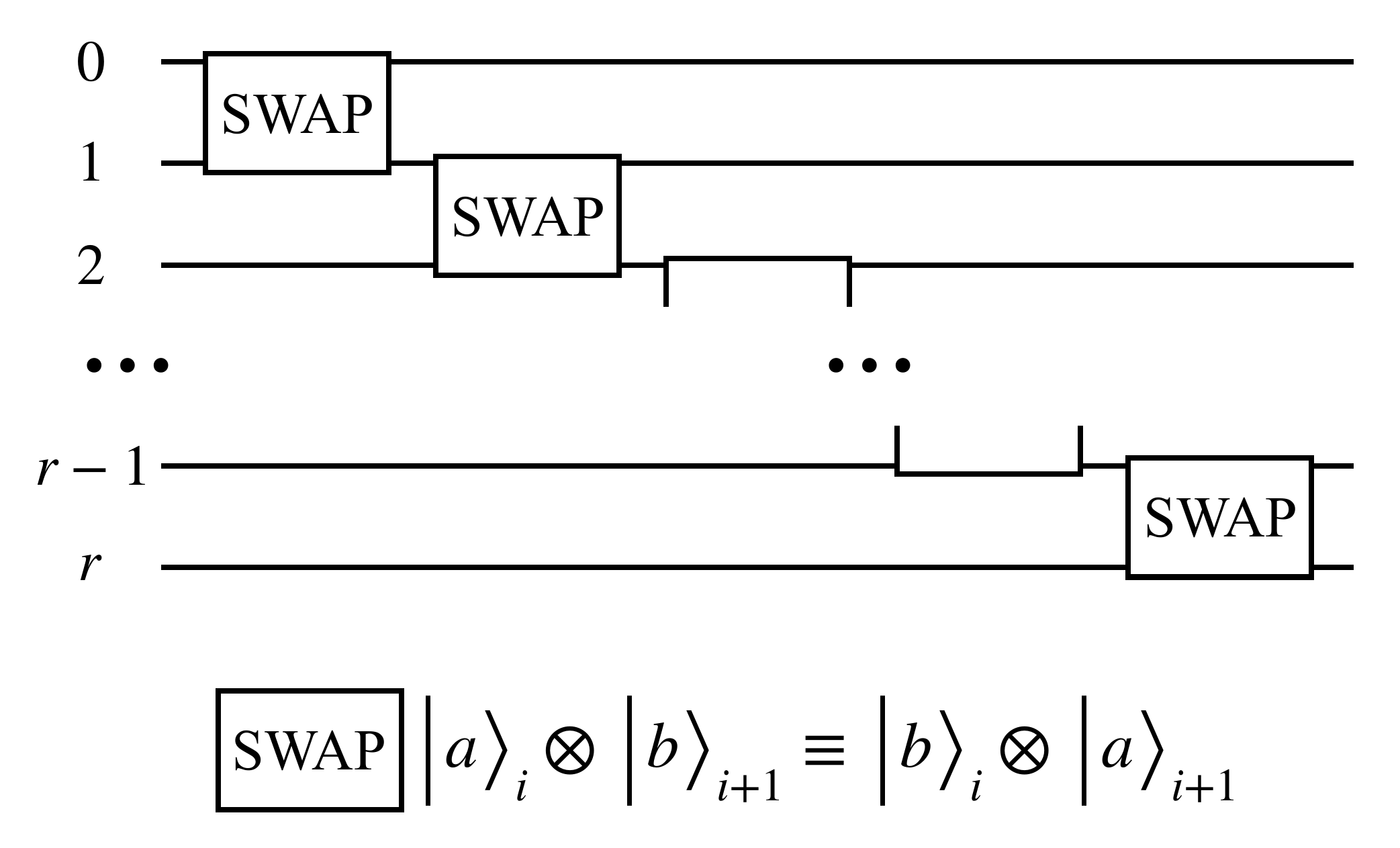}
\caption{Circuit diagram of a simple protocol that transfers the state on site 0 to site $r$ (here sites are arranged vertically and solid lines indicate the absence of any unitary). The definition of the SWAP gate acting on sites $i$ and $i+1$ is given at the bottom of the figure --- $a$ and $b$ label basis states of the local Hilbert space (each taking values in $\{1, \cdots, d\}$).}
\label{fig:transfer_protocol}
\end{figure}

We next derive more explicit bounds by considering the ``empirical distribution'' $\mu_r(J)$, defined (for a given set $\{J_l\}$) as the fraction of links between sites 0 and $r$ having $J_l \leq J$:
\begin{equation} \label{eq:empirical_distribution_definition}
\mu_r(J) \equiv \frac{1}{r} \sum_{l=1}^r \delta_{J_l \leq J},
\end{equation}
where $\delta_{J_l \leq J} \equiv 1$ if $J_l \leq J$ and 0 otherwise.
We assume that $\mu_r(J)$ converges as $r \rightarrow \infty$ (in a sense defined in Sec.~\ref{subsec:explicit_distribution_results}) to a function $\mu(J)$, and that the latter behaves as a power law at small $J$: $\mu(J) \sim \mu_0 J^{\alpha}$ with $\alpha > 0$.
The exponent $\alpha$ characterizes the prevalence of weak links, with smaller $\alpha$ implying more weak links.
Our analysis in fact applies for more general forms of $\mu(J)$ (as we discuss in Sec.~\ref{subsec:explicit_distribution_results}), but the power-law behavior is particularly convenient and representative.
Note that the convergence of $\mu_r(J)$ to $\mu(J)$  does not assume anything regarding the arrangement of weak links in space --- many of our results will hold regardless of where the weak links are located.

An essential feature of an LR bound is the shape of the ``front'', i.e., the spacetime curve $t(r)$ that separates the region in which the bound is small from that in which the bound is large (and thus vacuous).
The dynamical exponent $z$ and generalized LR velocity $v$ are defined by the asymptotic behavior $vt(r) \sim r^z$ at large $r$.
Keep in mind that $v$ has units of a genuine velocity only when $z = 1$ --- we will stick to the term ``generalized velocity''\footnote{An alternate quantity is $v_{\textrm{LR}}(r) \equiv [\textrm{d}t(r)/\textrm{d}r]^{-1}$~\cite{FossFeig2015Nearly}. Although distance-dependent, $v_{\textrm{LR}}(r)$ has the correct units of velocity even when $z \neq 1$.} for $z \neq 1$.
Whereas the conventional LR bound, Eq.~\eqref{eq:conventional_LR_bound}, has a ballistic front ($z = 1$) for all $\alpha > 0$, we show that the improved bound, Eq.~\eqref{eq:improved_LR_bound}, instead has
\begin{equation} \label{eq:threshold_dynamical_exponent_value}
z_c(\alpha) = \max \big[ \alpha^{-1}, 1 \big].
\end{equation}
The curve $z_c(\alpha)$ is sketched in Fig.~\ref{fig:main_result}.

One consequence of the above is that there cannot exist any $H \in \mathcal{H}_J$ for which the operator-spreading front grows with a dynamical exponent $z < z_c(\alpha)$.
In particular, it is impossible to have a ballistic front if $\alpha < 1$.
We have shaded this region blue in Fig.~\ref{fig:main_result} and labelled it as ``asymptotically unattainable on any site'' --- it is impossible to construct a Hamiltonian having that value of $\alpha$ which, at large distances, reaches any site at that value of $z$.

On the other hand, we also identify an $H \in \mathcal{H}_J$ whose front grows faster than any $z > z_c(\alpha)$, again requiring only that $\mu_r(J) \rightarrow \mu(J)$ in a suitably strong sense.
The Hamiltonian is rather straightforward, consisting simply of a sequential series of SWAP gates as shown in Fig.~\ref{fig:transfer_protocol}.
Once all gates have been applied, $A_0^t$ is supported on site $r$ and thus will generically fail to commute (by an $r$-independent amount) with $B_r$.
In order to satisfy $\lVert H_l(t) \rVert \leq J_l$, the total runtime of the circuit is proportional to $\sum_{l=1}^r 1/J_l$ --- analysis of this sum gives the behavior of the front.

We have shaded the region $z > z_c(\alpha)$ red in Fig.~\ref{fig:main_result} and labelled it as ``asymptotically attainable on all sites'' --- it is possible (and we do so) to construct a Hamiltonian having that value of $\alpha$ which reaches every large-distance site at that value of $z$.
In this sense, Eq.~\eqref{eq:threshold_dynamical_exponent_value} is \textit{the} dynamical exponent for a given $\alpha$, i.e., Eq.~\eqref{eq:threshold_dynamical_exponent_value} is tight.

It is rather striking that our result for $z_c(\alpha)$ agrees exactly with the value predicted on physical grounds in Ref.~\cite{Nahum2018Dynamics}.
The results of Ref.~\cite{Nahum2018Dynamics} are based on a coarse-grained description of 1D disordered systems, in which it is postulated that a region $l$ can be characterized by an effective ``growth rate'' $\Gamma_l$ setting the rate at which operators spread across the region.
The authors assume that $\Gamma_l$ is power-law-distributed with exponent $\alpha$ (although they work with the probability \textit{density} having exponent $\alpha - 1$), and ultimately deduce precisely Eq.~\eqref{eq:threshold_dynamical_exponent_value}.
Our results are somewhat more limited in scope\footnote{\label{note:phenomenology_comparison}In a different sense, our results are actually more general than those of Ref.~\cite{Nahum2018Dynamics} --- we derive bounds for an \textit{individual} realization of couplings, and thus are agnostic to its origin. For example, many of our results (namely those in Sec.~\ref{sec:general_bound}) hold equally well for quasiperiodic systems as for those with quenched randomness.}, since we consider \textit{microscopic} weak links rather than effective weak links emerging at long wavelengths, but this nonetheless provides a rigorous foundation for many of the concepts at work in Ref.~\cite{Nahum2018Dynamics}.
It would be of great interest to investigate whether the other phenomena discussed in Ref.~\cite{Nahum2018Dynamics}, such as entanglement growth and transport, can be placed on similar rigorous grounds.

Returning to Fig.~\ref{fig:main_result}, the situation becomes much more complicated on the boundary $z = z_c(\alpha)$, i.e., when considering dynamics on the scale set by $z_c(\alpha)$, both for our LR bounds and for our example Hamiltonians.
We demonstrate this by a detailed analysis of the special case in which the couplings are drawn independently from a literal probability distribution $\mu(J)$.
After proving that $\mu_r(J) \rightarrow \mu(J)$ in the required sense with probability 1, and thus that the portion of Fig.~\ref{fig:main_result} away from $z_c(\alpha)$ does indeed hold, we find that the generic behavior on the boundary cannot be described by a single bound --- we must introduce (at least) \textit{two types} of LR bounds:
\begin{itemize}
\item ``Almost-always'' (a.a.) bounds are those that hold for all sites $r$, excepting at most a finite number of sites.
In other words, there exists a distance $R$ such that the bound holds for all $r > R$.
\item ``Infinitely-often'' (i.o.) bounds are those that hold for an infinite subsequence of sites $\{r_k\}$ but need not hold outside of those sites.
In other words, for any distance $R$ there exists some site $r > R$ which is subject to the bound.
\end{itemize}

One can imagine situations in which either of the above two bounds is more relevant.
For example, suppose that Alice is manipulating one site of a spin chain and wants to be confident that her actions do not disturb distant regions in a certain amount of time.
In this case, a.a.\ bounds provide the desired guarantee.
On the other hand, suppose that Bob intends to transmit a signal along the spin chain.
If it is important that his signal reach every site faster than a certain rate, then i.o.\ bounds place the heaviest restrictions on what can be achieved.

In our case (still at $z = z_c(\alpha)$ and still assuming independent random couplings), we find different behaviors depending on how $\alpha$ compares to 1.
If $\alpha > 1$, the results are straightforward: our example Hamiltonian has a ballistic front that spreads to every site with finite velocity (note that $z_c(\alpha) = 1$ for $\alpha > 1$).
We have included this portion of the boundary with the red region in Fig.~\ref{fig:main_result} to indicate that it is also ``asymptotically attainable on all sites'' (albeit with a maximum allowed velocity).

However, we show that if $\alpha < 1$, then an i.o.\ bound having $z = z_c(\alpha)$ and arbitrarily small (generalized) velocity holds, while concurrently, our example Hamiltonian does reach a subsequence $\{r_k\}$ asymptotically faster than $z_c(\alpha)$ (hence no such a.a.\ bound can hold).
Both statements hold with probability 1.
Thus it is impossible to have a front which reaches \textit{every} site at dynamical exponent $z_c(\alpha)$, but it is possible (and we do so) to construct a Hamiltonian which reaches a \textit{subsequence} of sites at $z_c(\alpha)$. 
We have drawn this portion of the boundary purple in Fig.~\ref{fig:main_result} and accordingly labelled it ``asymptotically attainable on some but not all sites''.
Keep in mind that we have proven this final statement only for the special case of independent random couplings.
Nonetheless, it is a highly non-trivial example that makes clear the importance of distinguishing between a.a.\ and i.o.\ bounds.

Interestingly, the lone point $(\alpha, z) = (1, 1)$ is the only portion of Fig.~\ref{fig:main_result} in which we have been unable to give a definite answer.
An i.o.\ bound with vanishing velocity still holds, but the front in our example Hamiltonian is now sub-ballistic for every site.
It may be that a more complicated Hamiltonian exists which does reach a subsequence $\{r_k\}$ at finite velocity, yet it may instead be that a more sophisticated mathematical technique can produce an a.a.\ bound with vanishing velocity.
Further investigation is clearly warranted.

Lastly, we discuss the implications of our results for various applications.
The LR bounds themselves have physical content --- the statement derived here that ballistic spreading is impossible for $\alpha < 1$ can be considered an application in and of itself.
That said, our results have broader consequences as well.
Applications can roughly be grouped into two classes: those that follow from the existence of the front, and those that follow from the ``tail'' (i.e., the rapid decay of the LR bound at large distances outside the front).
Since we obtain a significantly altered front for $\alpha < 1$, our results have a qualitative impact on the former class.
However, while we do find a more complicated tail than in the conventional bound, the behavior at the largest distances turns out to be unmodified, and thus our results have only a minor impact on the latter class.

In the remainder of the paper, we make precise and prove the above statements.
Sec.~\ref{sec:definitions} establishes notation and the formalism within which we work.
Sec.~\ref{sec:general_bound}, after reviewing the conventional LR bound, derives Eq.~\eqref{eq:improved_LR_bound} for generic non-translation-invariant systems, and then makes use of the empirical distribution $\mu_r(J)$ to derive Fig.~\ref{fig:main_result}.
Sec.~\ref{sec:disordered_bound} considers the case of independent random couplings $\{J_l\}$ in more detail, first proving that the requirements of the preceding section are met and then examining behavior on the boundary $z_c(\alpha)$, with particular focus on the distinction between a.a.\ and i.o.\ behavior.
Sec.~\ref{sec:applications} lastly discusses the consequences of the above for various applications.

\section{Definitions \& notation} \label{sec:definitions}

\subsection{Geometry} \label{subsec:definition_geometry}

In this work, we consider an $N$-site lattice in 1D, where each site hosts a $d$-state degree of freedom.
In other words, the Hilbert space is a tensor product of $N$ local $d$-dimensional Hilbert spaces.
Let $\Omega$ denote the set of all $N$ sites and $\Lambda$ denote the set of all $N-1$ links.
Here we consider only \textit{nearest-neighbor} Hamiltonians on this lattice, i.e., Hamiltonians of the form $H(t) = \sum_{l \in \Lambda} H_l(t)$, where $H_l(t)$ is supported only on the sites connected by link $l$ (although as noted above in footnote~\ref{note:local_terms}, our results hold for Hamiltonians with arbitrary local terms as well).
These features are illustrated in Fig.~\ref{fig:lattice_layout}.

Given a set of couplings $\{J_l\}_{l \in \Lambda}$, let $\mathcal{H}_J$ be the family of all nearest-neighbor Hamiltonians for which
\begin{equation} \label{eq:constraint_definition}
\big\lVert H_l(t) \big\rVert \leq J_l.
\end{equation}
We even allow for a non-vanishing fraction of $\{J_l\}$ to be infinite, meaning that there is no restriction on the corresponding terms.
Pick an operator of interest $A_0$ supported only\footnote{\label{note:many_many}In fact, it will be clear from our proof technique that the support of $A_0$ need not be solely site 0 but can include all sites to the left of 0. Similarly, the support of $B_r$ can include all sites to the right of $r$. In other words, our bounds are ``many-to-many''.} on site 0, and similarly $B_r$ on site $r$ (the sites can be either well inside the chain or near the edges --- our bounds apply regardless).
For any $H \in \mathcal{H}_J$, let $A_0^t$ be the time evolution of $A_0$, i.e., the solution to\footnote{\label{note:Heisenberg}Strictly speaking, Eq.~\eqref{eq:operator_time_evolution_definition_v1} is \textit{not} the Heisenberg equation of motion for an arbitrary time-dependent Hamiltonian $H(t)$. The Heisenberg-picture operator $A_0^t$ is defined by the relation $\langle \Psi | A_0^t | \Psi \rangle \equiv \langle \Psi^t | A_0 | \Psi^t \rangle$, where $|\Psi^t \rangle$ is the solution to the Schrodinger equation: $i \partial_t |\Psi^t \rangle = H(t) |\Psi^t \rangle$ with $|\Psi^0 \rangle = |\Psi \rangle$. The solution can be written $|\Psi^t \rangle = \mathcal{T} e^{-i \int \textrm{d}s H(s)} |\Psi \rangle$, where the operator on the right-hand side is a time-ordered exponential --- earlier times appear to the right. Thus in the expression $\langle \Psi^t | A_0 | \Psi^t \rangle$ which defines $A_0^t$, earlier times appear \textit{outside} later times. The differential equation that correctly has $A_0^t$ as its solution is $\partial_s A_0^s = i[H(t-s), A_0^s]$ with $A_0^0 = A_0$ --- the commutator must be applied with later times first and earlier times last. Despite all of this, however, our analysis applies equally well with $H(t-s)$ in place of $H(s)$. Since the latter is less burdensome notation-wise, we shall use Eq.~\eqref{eq:operator_time_evolution_definition_v1} without further comment.}
\begin{equation} \label{eq:operator_time_evolution_definition_v1}
\partial_t A_0^t = i \big[ H(t), A_0^t \big], \qquad A_0^0 = A_0.
\end{equation}
Even though $A_0$ is supported on site 0, $A_0^t$ will (barring trivial cases) be supported throughout the entire chain for any $t > 0$.
The purpose of LR bounds is to place a bound on the quantity
\begin{equation} \label{eq:commutator_norm_definition}
D(r, t) \equiv \Big\lVert \big[ A_0^t, B_r \big] \Big\rVert,
\end{equation}
\textit{that holds uniformly for all $H \in \mathcal{H}_J$}.
Since only the ``portion'' of $A_0^t$ which acts non-trivially on site $r$ can possibly fail to commute with $B_r$ (see Eq.~\eqref{eq:general_trivial_bound} below), LR bounds considered as functions of $r$ and $t$ constrain the extent to which local operators ``spread'' throughout the system.

It is important to note that the Hamiltonians in $\mathcal{H}_J$ can have arbitrary time dependence, as long as Eq.~\eqref{eq:constraint_definition} is obeyed at all times.
Thus it is perhaps more informative to refer to any individual $H \in \mathcal{H}_J$ as a ``protocol'', since it can for example be a quantum circuit designed to perform a specific task.
As discussed in Sec.~\ref{sec:summary}, this distinction sheds light on the limitations of LR bounds.

The choice to use the operator norm in Eq.~\eqref{eq:commutator_norm_definition} has long been standard, as it enters naturally in many applications~\cite{Lieb1972Finite,Hastings2006Spectral,Bravyi2006LiebRobinson,Abanin2015Exponentially}.
There are situations in which alternative norms, in particular the Frobenius norm defined as $\lVert O \rVert_2 \equiv [d^{-N} \textrm{Tr} O^{\dag} O]^{1/2}$, may be more relevant and might behave quite differently~\cite{Tran2020Hierarchy}.
However, the operator norm is itself an upper bound on a wide family of norms including Frobenius (see App.~\ref{app:matrix_norms}).
Furthermore, the transfer protocol shown in Fig.~\ref{fig:transfer_protocol} leads to a commutator $[A_0^t, B_r]$ which is $O(1)$ using any of these norms.
Thus we shall exclusively consider the operator norm in this work, and the bounds obtained are automatically tight (at least regarding the dynamical exponent) for the other norms as well.

\subsection{Basis strings} \label{subsec:definition_basis_strings}

The set of Hermitian operators acting on the Hilbert space is itself a real vector space, and thus we can express any operator as a linear combination of certain basis operators.
First consider a single site $i$ and pick a Hermitian basis $\{X_i^{(\nu_i)}\}$ ($\nu_i \in \{0, 1, \cdots, d^2-1\}$).
For the entire chain, we use the tensor product basis $\{X^{(\nu)}\}$:
\begin{equation} \label{eq:tensor_product_basis_definition}
X^{(\nu)} \equiv \bigotimes_{i=1}^N X_i^{(\nu_i)},
\end{equation}
where $\nu \equiv (\nu_1, \cdots, \nu_N)$.
We assume (without loss of generality) that $\{X_i^{(\nu_i)}\}$ is chosen to be orthonormal with respect to the trace product, meaning that the tensor product basis is orthonormal as well:
\begin{equation} \label{eq:basis_orthonormality}
\begin{aligned}
d^{-N} \textrm{Tr} X^{(\nu)} X^{(\nu')} &= \prod_{i=1}^N d^{-1} \textrm{Tr} X_i^{(\nu_i)} X_i^{(\nu'_i)} \\
&= \prod_{i=1}^N \delta_{\nu_i \nu'_i} = \delta_{\nu \nu'}.
\end{aligned}
\end{equation}
We also take $X_i^{(0)}$ to be the identity $I_i$.
Beyond this, any choice of basis will work equally well.

We shall often refer to the basis elements as ``strings'', and define the support of a string to be the set of sites on which it does not have the identity:
\begin{equation} \label{eq:string_support_definition}
\textrm{supp} X^{(\nu)} \equiv \big\{ i: \nu_i \neq 0 \big\}.
\end{equation}
An important superoperator acting on the space of Hermitian operators is that which projects onto basis strings whose supports contain site $i$, i.e., strings that act non-trivially on site $i$.
We denote this superoperator by $\mathcal{P}_i$:
\begin{equation} \label{eq:projector_superoperator_definition}
\mathcal{P}_i X^{(\nu)} \equiv \Big( 1 - \delta_{\nu_i 0} \Big) X^{(\nu)}.
\end{equation}
Similarly, for any subset of sites $\omega \subseteq \Omega$, we define $\mathcal{P}_{\omega}$ to project onto basis strings which act non-trivially somewhere (not necessarily everywhere) within $\omega$.
A useful inequality (see App.~\ref{app:matrix_norms}) is that for any $\omega$ and any operator $O$,
\begin{equation} \label{eq:projected_operator_norm_bound}
\big\lVert \mathcal{P}_{\omega} O \big\rVert \leq 2 \big\lVert O \big\rVert.
\end{equation}
Also note that $[A_0^t, B_r] = [\mathcal{P}_r A_0^t, B_r]$, and so for $D(r, t)$ in Eq.~\eqref{eq:commutator_norm_definition}, we have the trivial bound
\begin{equation} \label{eq:general_trivial_bound}
D(r, t) \leq 2 \big\lVert \mathcal{P}_r A_0^t \big\rVert \big\lVert B_r \big\rVert.
\end{equation}
In what follows, we shall focus on bounding $\lVert \mathcal{P}_r A_0^t \rVert$, with a bound on $D(r, t)$ following automatically by Eq.~\eqref{eq:general_trivial_bound}.

The next important superoperator is the generator of time evolution under $H(t)$:
\begin{equation} \label{eq:evolution_generator_definition}
\mathcal{L}(t) O \equiv i \big[ H(t), O \big],
\end{equation}
and so (Eq.~\eqref{eq:operator_time_evolution_definition_v1})
\begin{equation} \label{eq:operator_time_evolution_definition_v2}
\partial_t A_0^t = \mathcal{L}(t) A_0^t.
\end{equation}
We will also need the generator corresponding to a subset of terms in the Hamiltonian.
For any subset of links $\lambda \subseteq \Lambda$, define
\begin{equation} \label{eq:subset_evolution_generator_definition}
\mathcal{L}_{\lambda}(t) O \equiv i \sum_{l \in \lambda} \big[ H_l(t), O \big].
\end{equation}
Clearly $\mathcal{L}(t) = \sum_{l \in \Lambda} \mathcal{L}_l(t)$.

Denote the evolution superoperator itself by $\mathcal{U}(t)$, i.e., $A_0^t \equiv \mathcal{U}(t) A_0$ (and define $\mathcal{U}_{\lambda}(t)$ analogously).
We can express the action of $\mathcal{U}(t)$ (and $\mathcal{U}_{\lambda}(t)$) in terms of a time-ordered exponential:
\begin{equation} \label{eq:evolution_superoperator_formal_expression}
\begin{aligned}
\mathcal{U}(t) A_0 &= \Big( \mathcal{T} e^{\int_0^t \textrm{d}s \mathcal{L}(s)} \Big) A_0 \\
&= \Big( \mathcal{T} e^{i \int_0^t \textrm{d}s H(s)} \Big) A_0 \Big( \mathcal{T} e^{i \int_0^t \textrm{d}s H(s)} \Big)^{\dag},
\end{aligned}
\end{equation}
where $\mathcal{T}$ denotes time-ordering (note the ordering in the bottom line of Eq.~\eqref{eq:evolution_superoperator_formal_expression} --- earlier times appear \textit{inside} later times).
Note that $\lVert \mathcal{U}(t) A_0 \rVert = \lVert A_0 \rVert$.

\begin{figure}[t]
\centering
\includegraphics[width=1.0\columnwidth]{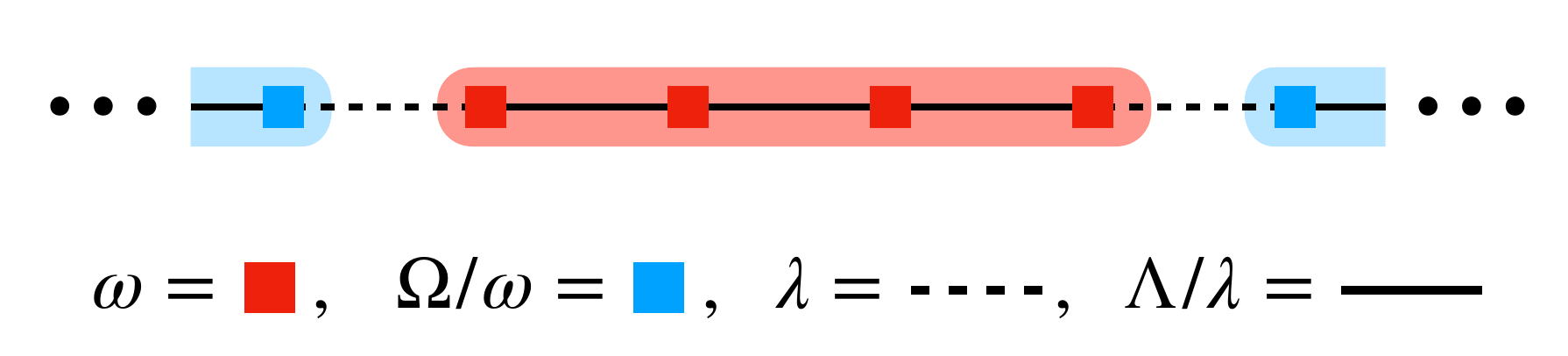}
\caption{Visual interpretation of Eq.~\eqref{eq:disconnected_lattice_commutation}. $\omega$ consists of the sites in red, $\Omega / \omega$ the sites in blue. $\lambda$ consists of the dashed links, $\Lambda / \lambda$ the solid links. Evolution under only $H_{\Lambda / \lambda}$ cannot transform a string having the identity on $\omega$ into one with a non-identity element, nor vice-versa.}
\label{fig:subset_notation}
\end{figure}

Lastly, suppose $\lambda \subseteq \Lambda$ contains every link which connects a subset of sites $\omega$ to its complement $\Omega / \omega$ (see Fig.~\ref{fig:subset_notation}).
It is intuitively clear that evolution under $\mathcal{L}_{\Lambda / \lambda}(t)$ alone cannot transform a basis operator which acts trivially on $\omega$ into one which acts non-trivially, or vice versa.
Put precisely,
\begin{equation} \label{eq:disconnected_lattice_commutation}
\mathcal{P}_{\omega} \mathcal{U}_{\Lambda / \lambda}(t) = \mathcal{U}_{\Lambda / \lambda}(t) \mathcal{P}_{\omega}.
\end{equation}
We give a proof of Eq.~\eqref{eq:disconnected_lattice_commutation} in App.~\ref{app:commutation_proof}.

\subsection{Types of Lieb-Robinson bounds} \label{subsec:definition_types}

To reiterate, the purpose of LR bounds is to place an upper limit on $\lVert \mathcal{P}_r A_0^t \rVert$ (and thus $D(r, t)$) which applies to every $H \in \mathcal{H}_J$ simultaneously. 
The bounds we construct will be of the form
\begin{equation} \label{eq:general_LR_bound_form}
\big\lVert \mathcal{P}_r A_0^t \big\rVert \leq \big\lVert A_0 \big\rVert f \left[ r^{\gamma} \left( 1 - \frac{vt}{r^z} \right) \right],
\end{equation}
where $\gamma > 0$ and the function $f(x)$ decays to zero as $x \rightarrow \infty$ and remains finite (or even diverges) as $x \rightarrow -\infty$.
A simple and common example is $f(x) = \exp{[-\kappa x]}$ for some $\kappa > 0$.
Although we shall not indicate so explicitly, note that all quantities here are functions of the couplings $\{J_l\}$ defining $\mathcal{H}_J$.

In the large-$r$ and large-$t$ limit, one identifies two important features from Eq.~\eqref{eq:general_LR_bound_form}:
\begin{itemize}
    \item There is a ``front'' defined by $vt = r^z$.
    For $vt > r^z$, the right-hand side of Eq.~\eqref{eq:general_LR_bound_form} is large (and thus the bound is vacuous), while for $vt < r^z$, $\lVert \mathcal{P}_r A_0(t) \rVert$ must be small.
    Thus the spacetime curve $vt = r^z$ constitutes an envelope that constrains the expansion of $A_0^t$.
    We refer to $z$ as the ``dynamical exponent'' and $v$ as the ``generalized velocity'' of the bound (only when $z = 1$ will we speak simply of the ``velocity'').
    Of particular interest are the \textit{largest} value of $z$ and \textit{smallest} value of $v$ for which a bound as in Eq.~\eqref{eq:general_LR_bound_form} holds.
    \item At fixed $t$, there is the ``tail'' behavior as $r \rightarrow \infty$, characterized by the exponent $\gamma$ and the large-$x$ behavior of $f(x)$ (e.g., exponential or power-law).
    Even though $\lVert \mathcal{P}_r A_0^t \rVert$ need never be identically zero, the tail describes how rapidly it must decay at large distances.
\end{itemize}

\begin{figure}[t]
\centering
\includegraphics[width=0.9\columnwidth]{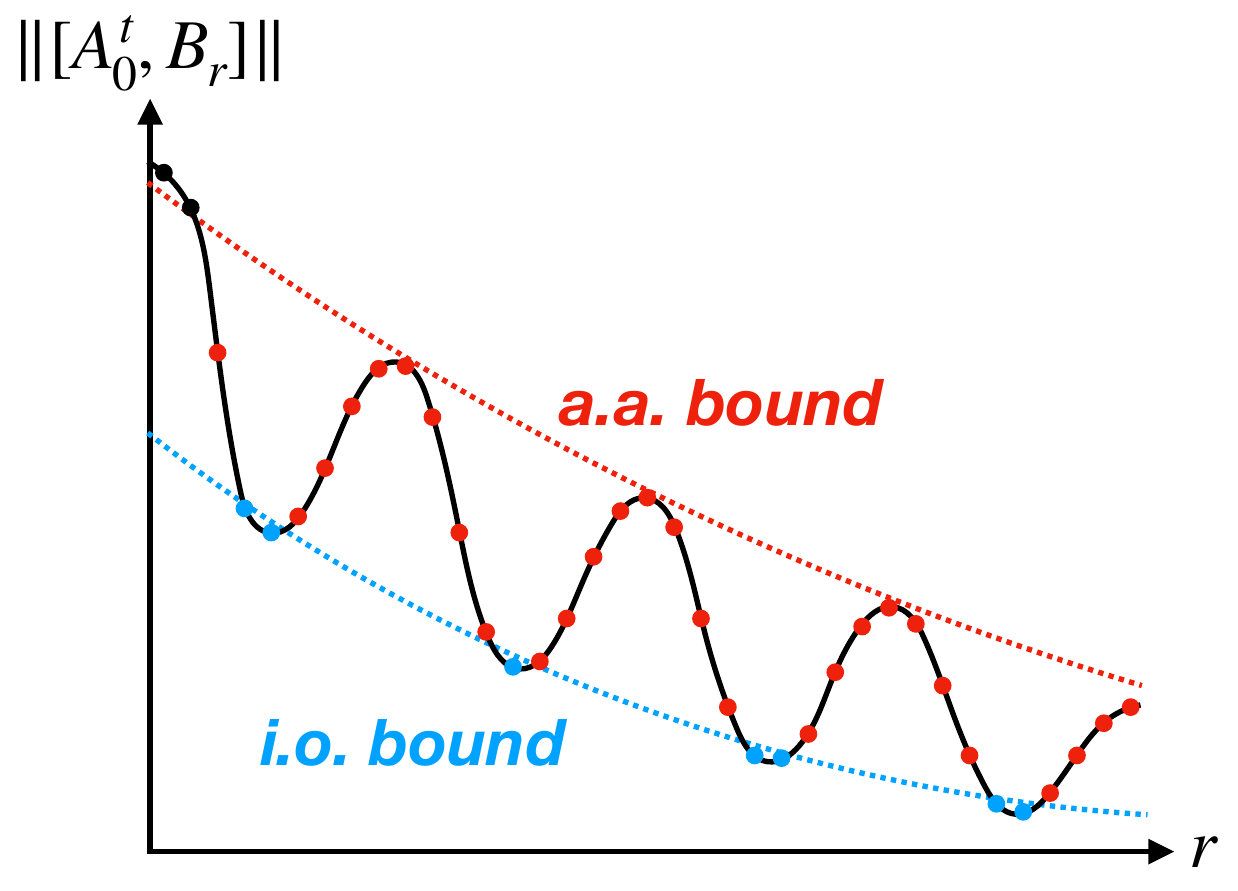}
\caption{Sketch of a possible spatial profile $\lVert [A_0^t, B_r] \rVert$ as a function of $r$, indicated by dots (black line merely connects the points). Red dashed line shows an a.a.\ bound consistent with the profile --- red and blue dots indicate points contained within the bound. There are implied to be only a finite number of black dots. Blue dashed line shows an i.o.\ bound ---  blue dots indicate the subsequence $\{r_k\}$ to which the bound applies.}
\label{fig:aa_io_distinction}
\end{figure}

As mentioned in Sec.~\ref{sec:summary}, we shall find it necessary to distinguish between different types of LR bounds, based on whether a statement such as Eq.~\eqref{eq:general_LR_bound_form} holds for \textit{all} sites or merely a \textit{subsequence} of sites:
\begin{itemize}
\item We call Eq.~\eqref{eq:general_LR_bound_form} an ``almost-always'' (a.a.) bound if there exists an $R$ such that it holds for all $r > R$.
This is the sense in which all past works (to our knowledge) have derived and discussed LR bounds\footnote{\label{note:almost_always_distinction}Strictly speaking, past works have derived bounds that hold for every single site (i.e., $R = 0$). Here we do not differentiate between ``almost-always'' and such ``always'' bounds (although one arguably could), since our focus is on asymptotic behavior. Given an a.a.\ bound with $R > 0$, one simple way to extend it to all sites is to use the trivial bound $\lVert [A_0^t, B_r] \rVert \leq 2 \lVert A_0 \rVert \lVert B_r \rVert$ for $r \leq R$. If the couplings $\{J_l\}$ are bounded by some $J_{\textrm{max}}$, another way is to use the conventional LR bound corresponding to $J_{\textrm{max}}$ for $r \leq R$ (Eq.~\eqref{eq:uniform_LR_bound} below). Since the number of sites which violate the a.a.\ bound is by definition finite, neither prescription changes any of the asymptotic behavior.}.
\item We call Eq.~\eqref{eq:general_LR_bound_form} an ``infinitely-often'' (i.o.) bound if for every $R$, it holds for some $r > R$.
This is equivalent to saying that there exists a subsequence $\{r_k\}_{k=1}^{\infty}$ on which the bound holds.
\end{itemize}
Fig.~\ref{fig:aa_io_distinction} illustrates the distinction between the two, showing a hypothetical curve $\lVert [A_0^t, B_r] \rVert$ versus $r$ alongside consistent a.a.\ and i.o.\ bounds.
Note that a.a.\ bounds are automatically i.o.\ bounds, but not vice-versa.
The distinction is likely unimportant for translation-invariant systems (although we are not aware of any works that compare the two to begin with), but it turns out to be essential in non-translation-invariant systems --- as mentioned in Sec.\ \ref{sec:summary}, there will be situations in which we can derive i.o.\ bounds but are unable to derive corresponding a.a.\ bounds (even proving that no such a.a.\ bound can exist).

\section{General bound for non-translation-invariant systems} \label{sec:general_bound}

\subsection{Review of the conventional bound} \label{subsec:review_old_bound}

We first review the standard Lieb-Robinson bound.
Start with a Hamiltonian $H \in \mathcal{H}_J$:
\begin{equation} \label{eq:general_Hamiltonian_form}
H(t) = \sum_{l \in \Lambda} H_l(t), \qquad \big\lVert H_l(t) \big\rVert \leq J_l.
\end{equation}
For any link $l$, pass to the interaction picture with respect to all other terms in the Hamiltonian by defining
\begin{equation} \label{eq:standard_interaction_picture_definition}
A_{0I}^t \equiv \mathcal{U}_{\Lambda / l}(t)^{\dag} A_0^t.
\end{equation}
The equation of motion for $A_{0I}^t$ (see Eqs.~\eqref{eq:operator_time_evolution_definition_v2} and~\eqref{eq:evolution_superoperator_formal_expression}) is
\begin{equation} \label{eq:standard_interaction_picture_EOM}
\begin{aligned}
\partial_t A_{0I}^t &= -\mathcal{U}_{\Lambda / l}(t)^{\dag} \mathcal{L}_{\Lambda / l}(t) A_0^t + \mathcal{U}_{\Lambda / l}(t)^{\dag} \mathcal{L}(t) A_0^t \\
&= \mathcal{U}_{\Lambda / l}(t)^{\dag} \mathcal{L}_l(t) A_0^t,
\end{aligned}
\end{equation}
from which it follows that
\begin{equation} \label{eq:standard_interaction_picture_formal_solution}
A_{0I}^t = A_0 + \int_0^t \textrm{d}s \, \mathcal{U}_{\Lambda / l}(s)^{\dag} \mathcal{L}_l(s) A_0^s.
\end{equation}

Since we are considering a 1D chain, $l$ is the only link that connects the sites on its right (denoted $>l$) to the sites on its left (denoted $<l$).
We thus use Eq.~\eqref{eq:disconnected_lattice_commutation} to obtain
\begin{equation} \label{eq:standard_projected_operator_solution}
\mathcal{P}_{>l} A_{0I}^t = \mathcal{P}_{>l} A_0 + \int_0^t \textrm{d}s \, \mathcal{U}_{\Lambda / l}(s)^{\dag} \mathcal{P}_{>l} \mathcal{L}_l(s) A_0^s.
\end{equation}
Furthermore, $\mathcal{P}_{>l} A_{0I}^t = \mathcal{U}_{\Lambda / l}(t)^{\dag} \mathcal{P}_{>l} A_0^t$.
Taking the norm then gives
\begin{equation} \label{eq:standard_projected_norm_inequality}
\big\lVert \mathcal{P}_{>l} A_0^t \big\rVert \leq \big\lVert \mathcal{P}_{>l} A_0 \big\rVert + \int_0^t \textrm{d}s \big\lVert \mathcal{P}_{>l} \mathcal{L}_l(s) A_0^s \big\rVert.
\end{equation}
Recall that $\mathcal{L}_l(s) A_0^s \equiv i [H_l(s), A_0^s]$.
Since $H_l(s)$ is supported only on link $l$,
\begin{equation} \label{eq:standard_projector_insertion}
\mathcal{L}_l(s) A_0^s = \mathcal{L}_l(s) \mathcal{P}_{>l-1} A_0^s.
\end{equation}
Together with Eq.~\eqref{eq:projected_operator_norm_bound}, we thus have
\begin{equation} \label{eq:standard_chain_projection_bound}
\begin{aligned}
\big\lVert \mathcal{P}_{>l} \mathcal{L}_l(s) \mathcal{P}_{>l-1} A_0^s \big\rVert &\leq 2 \big\lVert \mathcal{L}_l(s) \mathcal{P}_{>l-1} A_0^s \big\rVert \\
&\leq 4J_l \big\lVert \mathcal{P}_{>l-1} A_0^s \big\rVert,
\end{aligned}
\end{equation}
and Eq.~\eqref{eq:standard_projected_norm_inequality} becomes
\begin{equation} \label{eq:standard_LR_recursion_v1}
\big\lVert \mathcal{P}_{>l} A_0^t \big\rVert \leq \big\lVert \mathcal{P}_{>l} A_0 \big\rVert + 4J_l \int_0^t \textrm{d}s \big\lVert \mathcal{P}_{>l-1} A_0^s \big\rVert.
\end{equation}
Taking $r > 0$ for concreteness and supposing that $l$ lies between sites 0 and $r$, $\lVert \mathcal{P}_{>l} A_0 \rVert = 0$ and we are left with
\begin{equation} \label{eq:standard_LR_recursion_v2}
\big\lVert \mathcal{P}_{>l} A_0^t \big\rVert \leq 4J_l \int_0^t \textrm{d}s \big\lVert \mathcal{P}_{>l-1} A_0^s \big\rVert.
\end{equation}

To obtain a closed-form bound on $\lVert \mathcal{P}_r A_0^t \rVert$, first note that $\lVert \mathcal{P}_r A_0^t \rVert = \lVert \mathcal{P}_r \mathcal{P}_{\geq r} A_0^t \rVert \leq 2 \lVert \mathcal{P}_{\geq r} A_0^t \rVert$ (Eq.~\eqref{eq:projected_operator_norm_bound}), then use Eq.~\eqref{eq:standard_LR_recursion_v2} iteratively: bound $\lVert \mathcal{P}_{\geq r} A_0^t \rVert$ in terms of $\lVert \mathcal{P}_{\geq r-1} A_0^s \rVert$, then bound the latter in terms of $\lVert \mathcal{P}_{\geq r-2} A_0^{s'} \rVert$, and so on until reaching the origin (at which point do not introduce $\mathcal{P}_{\geq 0}$ --- simply use that $\lVert A_0^s \rVert = \lVert A_0 \rVert$).
The result is
\begin{equation} \label{eq:standard_LR_bound}
\big\lVert \mathcal{P}_r A_0^t \big\rVert \leq 2 \big\lVert A_0 \big\rVert \left( \prod_{l=1}^r 4J_l \right) \frac{t^r}{r!}.
\end{equation}

From here, one usually takes all $J_l$ to equal a common value $J$.
This gives (using that $r! \geq (r/e)^r$)
\begin{equation} \label{eq:uniform_LR_bound}
\begin{aligned}
\big\lVert \mathcal{P}_r A_0^t \big\rVert &\leq 2 \big\lVert A_0 \big\rVert \exp{\left[ r \log{\frac{4eJt}{r}} \right]} \\
&\leq 2 \big\lVert A_0 \big\rVert \exp{\left[ -r \left( 1 - \frac{4eJt}{r} \right) \right]}.
\end{aligned}
\end{equation}
Eq.~\eqref{eq:uniform_LR_bound} is of the form in Eq.~\eqref{eq:general_LR_bound_form}, with $z = 1$, $\gamma = 1$, and $v = 4eJ$: we have a ballistic (a.a.) LR bound with exponential tail and velocity of order $J$.

One can certainly use Eq.~\eqref{eq:standard_LR_bound} for non-translation-invariant $\{J_l\}$ as well.
Yet it is easy to see that the result might be rather weak.
Suppose that at large $r$, the empirical distribution of couplings (Eq.~\eqref{eq:empirical_distribution_definition}) approaches a function $\mu(J)$ (in some sufficiently strong sense --- we are only reasoning schematically for the moment).
Then
\begin{equation} \label{eq:sum_log_constraints_behavior}
\begin{aligned}
\prod_{l=1}^r 4J_l &= \exp{\left[ \sum_{l=1}^r \log{4J_l} \right]} \sim \exp{\left[ r \int \textrm{d}\mu(J) \log{4J} \right]}.
\end{aligned}
\end{equation}
As long as $\log{J}$ is integrable with respect to $\mu(J)$, we still obtain a ballistic LR bound:
\begin{equation} \label{eq:varying_weak_LR_bound}
\big\lVert \mathcal{P}_r A_0^t \big\rVert \leq 2 \big\lVert A_0 \big\rVert \exp{\left[ -r \left( 1 - \frac{4e \overline{J} t}{r} \right) \right]},
\end{equation}
where $\log{4 \overline{J}} \equiv \int \textrm{d}\mu(J) \log{4J}$.
Yet the requirement that $\log{J}$ be integrable is quite easy to satisfy: if $\mu(J) \sim J^{\alpha}$ at small $J$ for \textit{any} $\alpha > 0$, however small, then $\overline{J}$ is finite and Eq.~\eqref{eq:standard_LR_bound} is ballistic.
Clearly this is a much weaker claim than that in Fig.~\ref{fig:main_result}.
As we show in the following subsections, taking the small-$J$ links more seriously allows us to prove that the actual dynamics must be sub-ballistic for any $\alpha < 1$.

\subsection{Improvements via integrating out links} \label{subsec:integrating_links}

\begin{figure}[t]
\centering
\includegraphics[width=1.0\columnwidth]{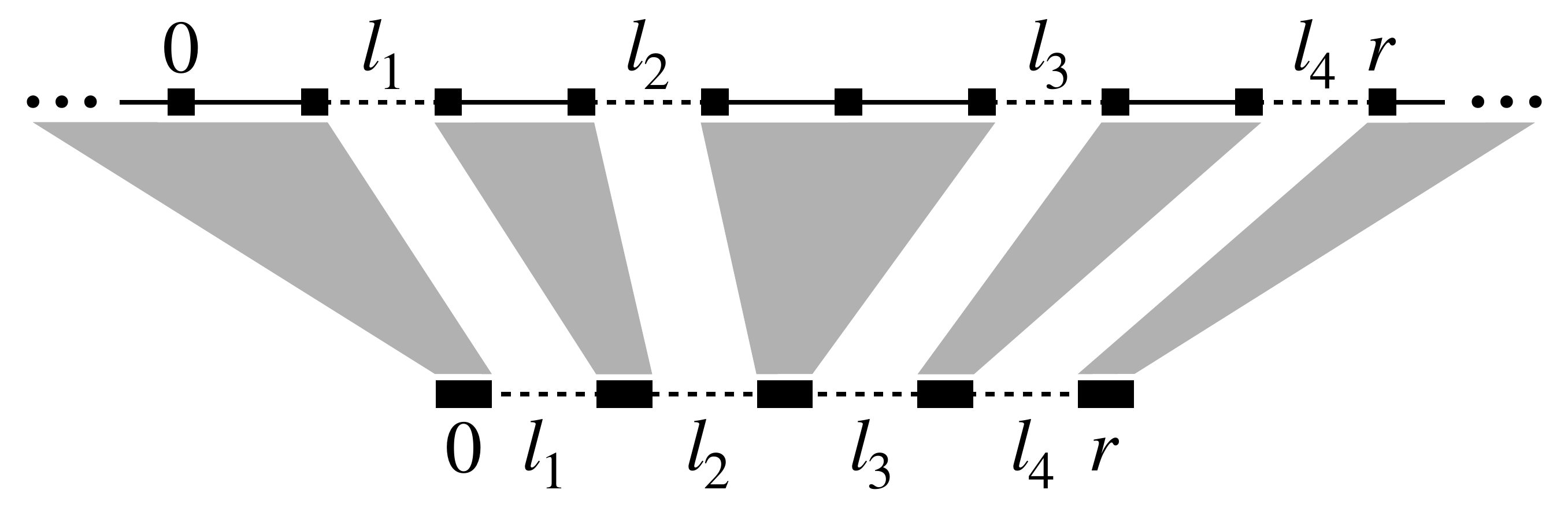}
\caption{Transformation from the original picture (top line) to the interaction picture with respect to ``large'' terms (bottom line). The dashed links (labelled $l_1$ through $l_4$) are those on which $\lVert H_l(t) \rVert \leq \epsilon$, and the solid links those on which $\lVert H_l(t) \rVert > \epsilon$. Squares on the top line represent individual sites, but rectangles on the bottom line represent the collection of all corresponding sites connected by solid links --- the transformed interaction $\widetilde{H}_{l_i}^t$ is supported on the entire neighboring rectangles (but no farther by virtue of Eq.~\eqref{eq:disconnected_lattice_commutation}).}
\label{fig:interaction_picture}
\end{figure}

In words, we improve on the conventional bound by passing to a further interaction picture with respect to all ``large'' terms in the Hamiltonian, namely all $H_l(t)$ whose norms exceed\footnote{\label{note:long_range}It is interesting to note that this same technique has been used to derive optimal LR bounds for power-law-interacting systems, e.g., $H(t) = \sum_{ij} H_{ij}(t)$ with $\lVert H_{ij}(t) \rVert \leq |i - j|^{-\alpha}$. To improve on the conventional bound, which would predict a front growing exponentially in time, group the interactions into short-range ($|i - j| \leq r^*$ for some $r^*$) and long-range ($|i - j| > r^*$) and then pass to the interaction picture with respect to short-range terms~\cite{FossFeig2015Nearly}. It has recently been shown that this procedure can be repeated for a hierarchy of length scales to obtain LR bounds which are provably tight~\cite{Tran2021LiebRobinson}.} some threshold $\epsilon$.
On the one hand, the remaining terms have a larger support in this interaction picture, and we do not try to describe their structure within that support.
Yet in return, \textit{every} remaining term has norm less than $\epsilon$, and so no dynamics can occur on any scale faster than $\epsilon^{-1}$.
The latter effect, which suppresses operator spreading, turns out to be the dominant one when there is a sufficient number of weak links.
We sketch the situation in Fig.~\ref{fig:interaction_picture}.

We now make this argument precise.
Pick any subset of links $\lambda \equiv \{l_1, \cdots, l_n\}$ lying between 0 and $r$.
Pass to the interaction picture with respect to $H_{\Lambda / \lambda}$:
\begin{equation} \label{eq:subset_interaction_picture_definition}
\widetilde{A}_0^t \equiv \mathcal{U}_{\Lambda / \lambda}(t)^{\dag} A_0^t, \qquad \widetilde{H}_l(t) \equiv \mathcal{U}_{\Lambda / \lambda}(t)^{\dag} H_l(t).
\end{equation}
As in Eq.~\eqref{eq:standard_interaction_picture_EOM}, the equation of motion for $\widetilde{A}_0^t$ is
\begin{equation} \label{eq:subset_interaction_picture_EOM}
\begin{aligned}
\partial_t \widetilde{A}_0^t &= \mathcal{U}_{\Lambda / \lambda}(t)^{\dag} \mathcal{L}_{\lambda}(t) \mathcal{U}_{\Lambda / \lambda}(t) \widetilde{A}_0^t \\
&\equiv \widetilde{\mathcal{L}}_{\lambda}(t) \widetilde{A}_0^t,
\end{aligned}
\end{equation}
where $\widetilde{\mathcal{L}}_{\lambda}(t) \widetilde{A}_0^t = i [\widetilde{H}_{\lambda}(t), \widetilde{A}_0^t]$ and $\widetilde{H}_{\lambda}(t) = \sum_{i=1}^n \widetilde{H}_{l_i}(t)$.

The transformed operator $\widetilde{H}_{l_i}(t)$ is supported on (potentially) all sites between $l_{i-1}$ and $l_{i+1}$, yet it has the same norm as $H_{l_i}(t)$, namely bounded by $J_{l_i}$.
Thus we can apply the same procedure as in Sec.~\ref{subsec:review_old_bound} to the operator $\widetilde{A}_0^t$, with $\lambda$ in place of $\Lambda$ and with $\widetilde{H}_{\lambda}(t)$ as the Hamiltonian.
Following an identical derivation to that of Eq.~\eqref{eq:standard_LR_recursion_v2}, we obtain
\begin{equation} \label{eq:subset_LR_recursion}
\big\lVert \mathcal{P}_{>l_i} \widetilde{A}_0^t \big\rVert \leq 4J_{l_i} \int_0^t \textrm{d}s \big\lVert \mathcal{P}_{>l_{i-1}} \widetilde{A}_0^s \big\rVert.
\end{equation}
Use that $\lVert \mathcal{P}_r A_0^t \rVert \leq 2 \lVert \mathcal{P}_{>l_n} A_0^t \rVert = 2 \lVert \mathcal{P}_{>l_n} \widetilde{A}_0^t \rVert$ to start the iteration, and $\lVert \widetilde{A}_0^s \rVert = \lVert A_0 \rVert$ to terminate it.
Thus
\begin{equation} \label{eq:subset_LR_bound}
\big\lVert \mathcal{P}_r A_0^t \big\rVert \leq 2 \big\lVert A_0 \big\rVert \left( \prod_{i=1}^n 4J_{l_i} \right) \frac{t^n}{n!}.
\end{equation}
Since $\lambda$ is arbitrary, we are free to use the subset that gives the tightest bound:
\begin{equation} \label{eq:general_improved_LR_bound}
\big\lVert \mathcal{P}_r A_0^t \big\rVert \leq 2 \big\lVert A_0 \big\rVert \min_{\lambda} \left[ \left( \prod_{i=1}^{|\lambda|} 4J_{l_i} \right) \frac{t^{|\lambda|}}{|\lambda|!} \right],
\end{equation}
where the minimum is over all subsets of links between 0 and $r$.
Eq.~\eqref{eq:general_improved_LR_bound} is our improved LR bound.

Although the minimization in Eq.~\eqref{eq:general_improved_LR_bound} may seem computationally expensive due to the $2^r$ possible $\lambda$, it can be performed efficiently.
For a fixed size of $\lambda$, the optimal choice is clearly those links having the $|\lambda|$ smallest values of $J_l$.
Thus one need only sort $\{J_l\}_{l=1}^r$ beforehand, and the minimization amounts to simply checking the $r$ possible values of $|\lambda|$.

\subsection{An explicit bound in terms of the distribution of couplings} \label{subsec:explicit_distribution_results}

We now make use of Eq.~\eqref{eq:general_improved_LR_bound} to prove general results in terms of the ``distribution'' of couplings $\mu_r(J)$, by which we mean the fraction of links between 0 and $r$ for which $J_l \leq J$.
Reproducing Eq.~\eqref{eq:empirical_distribution_definition} (recall that $\delta_{J_l \leq J} \equiv 1$ if $J_l \leq J$ and 0 otherwise),
\begin{equation} \label{eq:empirical_distribution_definition_repeat}
\mu_r(J) \equiv \frac{1}{r} \sum_{l=1}^r \delta_{J_l \leq J}.
\end{equation}
From this definition, for any $\epsilon > 0$ such that $\mu_r(\epsilon) \neq 0$, we can choose $\lambda$ to be the subset of links with $J_l \leq \epsilon$ and thus have the LR bound
\begin{equation} \label{eq:general_distribution_LR_bound}
\begin{aligned}
\big\lVert \mathcal{P}_r A_0^t \big\rVert &\leq 2 \big\lVert A_0 \big\rVert \frac{(4 \epsilon t)^{r \mu_r(\epsilon)}}{[r \mu_r(\epsilon)]!} \\
&\leq 2 \big\lVert A_0 \big\rVert \exp{\left[ -r \mu_r(\epsilon) \left( 1 - \frac{4e \epsilon t}{r \mu_r(\epsilon)} \right) \right]}.
\end{aligned}
\end{equation}

Now suppose that $\mu_r(J)$ converges\footnote{\label{note:infinite_limit}One might be concerned by us taking $r \rightarrow \infty$ since we are considering a finite chain of $N$ sites. However, note that all of the bounds derived in Secs.~\ref{sec:general_bound} and~\ref{sec:disordered_bound} are independent of $N$. Thus to be precise, we are supposing that we have an \textit{infinite} sequence of couplings $\{J_l\}$ which obeys Eqs.~\eqref{eq:empirical_distribution_convergence_1} and~\eqref{eq:empirical_distribution_convergence_2}, and we are considering the restriction of an infinite lattice to $N$ sites. Any statement involving $r$ should be interpreted as applying when $N > r$.} as $r \rightarrow \infty$ to a function $\mu(J)$ defined on $[0, \infty)$, with the small-$J$ (``tail'') behavior\footnote{\label{note:tilde}Throughout this paper, a statement such as ``$f(x) \sim g(x)$ as $x \rightarrow 0$'' means that $\lim_{x \rightarrow 0} f(x)/g(x) = 1$.}
\begin{equation} \label{eq:empirical_distribution_tail_behavior}
\mu(J) \sim \mu_0 J^{\alpha},
\end{equation}
for some $\mu_0 > 0$.
The parameter $\alpha > 0$ will play an essential role in what follows.
We do need to specify the precise sense in which $\mu_r(J)$ converges.
It will not be sufficient to require merely that $\lim_{r \rightarrow \infty} \mu_r(J) = \mu(J)$ at any fixed $J$, but also that $\mu_r(J)$ behave as $\mu(J)$ on scales which decrease with increasing $r$.
This is expressed by the following conditions, which we assume to be met:
\begin{itemize}
\item For any $\beta \in [0, 1/\alpha)$ and $J > 0$,
\begin{equation} \label{eq:empirical_distribution_convergence_1}
\lim_{r \rightarrow \infty} \frac{\mu_r(J r^{-\beta})}{\mu(J r^{-\beta})} = 1.
\end{equation}
\item For any $\beta > 1/\alpha$,
\begin{equation} \label{eq:empirical_distribution_convergence_2}
\lim_{r \rightarrow \infty} \frac{\min_{l=1}^r J_l}{r^{-\beta}} = \infty.
\end{equation}
\end{itemize}

As an example of why Eqs.~\eqref{eq:empirical_distribution_convergence_1} and~\eqref{eq:empirical_distribution_convergence_2} are necessary, rather than simply the condition $\lim_{r \rightarrow \infty} \mu_r(J) = \mu(J)$ (which is contained in Eq.~\eqref{eq:empirical_distribution_convergence_1} as the case $\beta = 0$), suppose that there is a single link on which $J_l = 0$. Clearly there cannot be any operator spreading past link $l$ (note that this is captured by the general bound in Eq.~\eqref{eq:general_improved_LR_bound}). On the other hand, the value of a single coupling does not affect the \textit{fraction} which are less than any value in the large-$r$ limit. Thus $\lim_{r \rightarrow \infty} \mu_r(J)$ at any fixed value of $J$ does not identify \textit{individual} anomalously weak links, whereas Eq.~\eqref{eq:empirical_distribution_convergence_2} does.
In other words, Eqs.~\eqref{eq:empirical_distribution_convergence_1} and~\eqref{eq:empirical_distribution_convergence_2} are a precise way of stating that the \textit{weakest} links between sites 0 and $r$ are also distributed in a manner behaving as $\mu(J)$ at large $r$.

Strictly speaking, we only need (and will prove) Eqs.~\eqref{eq:empirical_distribution_convergence_1} and~\eqref{eq:empirical_distribution_convergence_2} to hold for a dense subset of $\beta$, say all rational $\beta$.
The exponent $1/\alpha$ appears because $r \mu(J r^{-1/\alpha}) = O(1)$ --- one expects to find couplings with $\beta < 1/\alpha$ but not with $\beta > 1/\alpha$ between sites 0 and $r$.

We feel that these conditions are reasonable to expect in practice.
We show in Sec.~\ref{sec:disordered_bound} that if each $J_l$ is chosen independently from a literal probability distribution $\mu(J)$ obeying Eq.~\eqref{eq:empirical_distribution_tail_behavior}, then Eqs.~\eqref{eq:empirical_distribution_convergence_1} and~\eqref{eq:empirical_distribution_convergence_2} are satisfied with probability 1 --- in this sense, \textit{any} sufficiently disordered set of couplings will meet our requirements.
Whether Eqs.~\eqref{eq:empirical_distribution_convergence_1} and~\eqref{eq:empirical_distribution_convergence_2} hold in any specific situation obviously depends on the system under consideration, but in any case (and perhaps more importantly), one can always return to Eq.~\eqref{eq:general_improved_LR_bound} if needed.

Let us first consider $\alpha \geq 1$.
From Eq.~\eqref{eq:empirical_distribution_tail_behavior} and the assumption that $\lim_{r \rightarrow \infty} \mu_r(J) = \mu(J)$, there exists\footnote{\label{note:defining_limit}Let us remind the reader of the precise definition of the limit of a sequence: the statement $\lim_{r \rightarrow \infty} x_r = x$ is the statement that for any $\eta > 0$, there exists $R(\eta) \in \mathbb{N}$ such that $|x_r - x| < \eta$ for all $r > R(\eta)$.} $R$ and $\epsilon > 0$ such that for all $r > R$, $\mu_r(\epsilon) > C \epsilon^{\alpha}$ (recall our convention that $C$ is any $r$- and $t$-independent constant whose value may change from line to line).
Thus Eq.~\eqref{eq:general_distribution_LR_bound} becomes (for $r > R$)
\begin{equation} \label{eq:large_alpha_tail_upper_bound}
\big\lVert \mathcal{P}_r A_0^t \big\rVert \leq 2 \big\lVert A_0 \big\rVert \exp{\left[ -C \epsilon^{\alpha} r \left( 1 - \frac{4et}{C \epsilon^{\alpha - 1} r} \right) \right]}.
\end{equation}
This is simply a conventional LR bound with ballistic front and exponential tail (a.a.\ because it holds for \textit{all} $r > R$), having velocity $4e/C \epsilon^{\alpha - 1}$.

Now consider $\alpha < 1$.
Pick any $\beta \in (0, 1/\alpha)$, and then from Eqs.~\eqref{eq:empirical_distribution_tail_behavior} and~\eqref{eq:empirical_distribution_convergence_1}, there exists $R$ such that for all $r > R$, $\mu_r(r^{-\beta}) > C r^{-\beta \alpha}$.
Taking $\epsilon = r^{-\beta}$ in Eq.~\eqref{eq:general_distribution_LR_bound}, we thus have that for all $r > R$,
\begin{equation} \label{eq:general_tail_upper_bound_v1}
\big\lVert \mathcal{P}_r A_0^t \big\rVert \leq 2 \big\lVert A_0 \big\rVert \exp{\left[ -C r^{1 - \beta \alpha} \left( 1 - \frac{4et}{C r^{1 + \beta - \beta \alpha}} \right) \right]}.
\end{equation}
This is an a.a.\ LR bound having exponents
\begin{equation} \label{eq:general_LR_dynamical_exponent}
z = 1 + \beta - \beta \alpha,
\end{equation}
\begin{equation} \label{eq:general_LR_tail_exponent}
\gamma = 1 - \beta \alpha.
\end{equation}
Note that the front is sub-ballistic precisely for $\alpha < 1$ (whereas it is no tighter than Eq.~\eqref{eq:large_alpha_tail_upper_bound} for $\alpha \geq 1$).

By taking $\beta \nearrow 1/\alpha$, we can make the dynamical exponent $z$ arbitrarily close to $1/\alpha$.
This actually implies that for any $z < 1/\alpha$, we have an a.a.\ LR bound \textit{with arbitrarily small generalized velocity}.
Fixing $z$ and for any $v > 0$, setting $\beta \in (z, 1/\alpha)$ and taking $r$ sufficiently large (so that $4er^{-\beta} < Cvr^{1 - \beta \alpha - z}$) gives
\begin{equation} \label{eq:general_tail_upper_bound_v2}
\begin{aligned}
\big\lVert \mathcal{P}_r A_0^t \big\rVert &\leq 2 \big\lVert A_0 \big\rVert \exp{\left[ -C r^{1 - \beta \alpha} \left( 1 - \frac{4et}{C r^{1 + \beta - \beta \alpha}} \right) \right]} \\
&\leq 2 \big\lVert A_0 \big\rVert \exp{\left[ -C r^{1 - \beta \alpha} \left( 1 - \frac{vt}{r^z} \right) \right]}.
\end{aligned}
\end{equation}
Interestingly, the tightest tail corresponds to the opposite limit of $\beta$.
Setting $\beta = 0$ gives the standard exponential tail (albeit only at distances $r > vt$ for some $v$), whereas increasing $\beta$ gives an increasingly stretched-exponential tail.
The optimal choice of $\beta$ depends on the specific application: one should take $\beta \nearrow 1/\alpha$ if constraining the shape of the front is most important, but one should set $\beta = 0$ if constraining the tail is most important.

We can combine these results (Eq.~\eqref{eq:large_alpha_tail_upper_bound} for $\alpha \geq 1$ and Eq.~\eqref{eq:general_tail_upper_bound_v2} for $\alpha < 1$) simply by saying that Eq.~\eqref{eq:general_tail_upper_bound_v2} holds for any $z < z_c(\alpha) \equiv \max[1/\alpha, 1]$.
This accounts for the blue region in Fig.~\ref{fig:main_result}, with the upper boundary being given by $z_c(\alpha)$.

Note that this analysis can straightforwardly be extended to limiting distributions $\mu(J)$ which are not simple power laws, with the expected results.
First of all, if $\mu(J)$ decays to 0 at small $J$ faster than any power law (e.g., $\mu(J) \sim \exp{[-1/J]}$), then a conventional LR bound as in Eq.~\eqref{eq:large_alpha_tail_upper_bound} still holds.
If $\mu(J)$ decays slower than any power law (e.g., $\mu(J) \sim 1/\log{J^{-1}}$), and if Eq.~\eqref{eq:empirical_distribution_convergence_1} is obeyed for all $\beta > 0$, then $z = \infty$ in that an LR bound with infinitesimal $v$ holds for any finite $z$.
Lastly, our main result still applies if $\mu(J)$ scales not solely as $J^{\alpha}$ but as $J^{\alpha} p(J)$ for some sub-power-law function $p(J)$ --- an LR bound with arbitrarily small generalized velocity holds for any $z < z_c(\alpha)$, as in Eq.~\eqref{eq:general_tail_upper_bound_v2}.

\subsection{Tightness of the bound} \label{subsec:bound_tightness}

As discussed in Sec.~\ref{sec:summary}, LR bounds should be complemented by an understanding of their tightness, ideally by constructing an explicit protocol $H \in \mathcal{H}_J$ that saturates the bound.
To that end, we consider the simple transfer protocol shown in Fig.~\ref{fig:transfer_protocol}.
Denoting the total runtime of the circuit by $T_r$, clearly $\mathcal{P}_r A_0^{T_r} = A_0$, and so any valid LR bound must have a front which encompasses the spacetime point $(r, T_r)$.
We shall focus on the dynamical exponent --- if $T_r = O(r^z)$, then no LR bound can have a dynamical exponent larger than $z$.

Effecting a SWAP gate for arbitrary $d$-dimensional local Hilbert spaces is not entirely trivial\footnote{\label{note:SWAP}As pointed out to us by P.\ J.\ D.\ Crowley and C.\ R.\ Laumann, it is not necessary to give an explicit construction of the SWAP gate. From the definition of SWAP, the quantity $H_l \equiv i\pi^{-1} J_l \log{\textrm{SWAP}_l}$ is a Hermitian operator supported on link $l$ with norm $J_l$. The Hamiltonian that applies $H_l$ sequentially as in Fig.~\ref{fig:transfer_protocol} for time $\pi J_l^{-1}$, although not given in a completely constructive form, is a legitimate member of $\mathcal{H}_J$.}, but a construction is given in Ref.~\cite{GarciaEscartin2013Swap}.
For completeness, we give the relevant details in App.~\ref{app:transfer_protocol}.
The only interactions needed (per SWAP gate) are a finite number of controlled-$Z$ gates.
In our case, since Eq.~\eqref{eq:constraint_definition} must be respected, the time per controlled-$Z$ gate across link $l$ is $O(1/J_l)$.
Thus the total runtime is
\begin{equation} \label{eq:transfer_runtime_expression}
T_r = C \sum_{l=1}^r \frac{1}{J_l}.
\end{equation}

We again assume that the distribution of couplings $\mu_r(J)$ satisfies Eqs.~\eqref{eq:empirical_distribution_convergence_1} and~\eqref{eq:empirical_distribution_convergence_2}.
As we demonstrate below, it then follows that for any $z > \max[1/\alpha, 1]$,
\begin{equation} \label{eq:transfer_runtime_result}
\lim_{r \rightarrow \infty} \frac{T_r}{r^z} = 0.
\end{equation}

Let us first note that it is the same threshold exponent $z_c(\alpha) \equiv \max[1/\alpha, 1]$ which enters into both Eq.~\eqref{eq:transfer_runtime_result} and~\eqref{eq:general_tail_upper_bound_v2}.
We can thus say that the dynamical exponent is $z_c(\alpha)$ in the following sense:
\begin{itemize}
\item For any $z < z_c(\alpha)$, there is no $H \in \mathcal{H}_J$ that can generate correlations at \textit{any} sufficiently large distance $r$ in a time of order $r^z$.
\item For any $z > z_c(\alpha)$, we know of an explicit protocol $H \in \mathcal{H}_J$ that can generate correlations at \textit{every} sufficiently large distance $r$ in a time vanishing compared to $r^z$.
\end{itemize}
However, the behavior precisely at $z = z_c(\alpha)$ is far more complicated and system-dependent.
In particular, the distinction between a.a.\ and i.o.\ bounds becomes essential, as we demonstrate explicitly in Sec.~\ref{sec:disordered_bound}.

Now we turn to the proof of Eq.~\eqref{eq:transfer_runtime_result}.
It will be convenient to define $Y_l \equiv 1/J_l$, so that $T_r = \sum_{l=1}^r Y_l$.
Note that $\mu_r(J)$, the fraction of links with $J_l \leq J$, is equivalently the fraction with $Y_l \geq 1/J$.
Fix $\gamma > 0$ and define $a_0 \equiv 0$, $a_k \equiv r^{(k-1) \gamma}$ for $k \geq 1$ (writing $a_k$ instead of $a_k(r)$ for conciseness).
Also define $p_k$ to be the fraction of links with $Y_l \in [a_k, a_{k+1})$, equivalently
\begin{equation} \label{eq:interval_fraction_definition}
p_k \equiv \mu_r \big( a_k^{-1} \big) - \mu_r \big( a_{k+1}^{-1} \big).
\end{equation}
By definition, we have the bound
\begin{equation} \label{eq:transfer_runtime_initial_bound}
T_r \leq r \sum_{k=0}^{\infty} p_k a_{k+1}.
\end{equation}

For $p_0$ and $p_1$, we shall simply use that $p_0 \leq 1$, $p_1 \leq 1$.
For $p_k$ with $k \geq 2$, first note that the second term in Eq.~\eqref{eq:interval_fraction_definition} can be neglected relative to the first at large $r$.
Thus it follows from Eqs.~\eqref{eq:empirical_distribution_tail_behavior} and~\eqref{eq:empirical_distribution_convergence_1} that for any $\eta > 0$, there exists $R_k$ such that for all $r > R_k$,
\begin{equation} \label{eq:interval_fraction_estimate}
p_k \in \big( (1 - \eta) C r^{-(k-1) \alpha \gamma}, (1 + \eta) C r^{-(k-1) \alpha \gamma} \big).
\end{equation}
Furthermore, Eq.~\eqref{eq:empirical_distribution_convergence_2} implies that if we take $K$ to be the smallest integer greater than $1/\alpha \gamma$, then there exists $R_{\infty}$ such that for all $r > R_{\infty}$, $\min_{l=1}^r J_l > r^{-K \gamma} = a_{K+1}^{-1}$ and therefore $p_k = 0$ for all $k \geq K+1$.
All together, we have that for $r > \max[R_2, \cdots, R_K, R_{\infty}]$,
\begin{equation} \label{eq:interval_fraction_summary}
p_k \leq \begin{cases} 1, \quad & k \leq 1 \\ (1 + \eta) C r^{-(k-1) \alpha \gamma}, \quad & 2 \leq k \leq K \\ 0, \quad & K+1 \leq k \end{cases}.
\end{equation}
Eq.~\eqref{eq:transfer_runtime_initial_bound} becomes
\begin{equation} \label{eq:transfer_runtime_bound_summation}
T_r \leq r + r^{1 + \gamma} + (1 + \eta) C r^{1 + \alpha \gamma} \sum_{k=2}^K r^{k(1 - \alpha) \gamma}.
\end{equation}

First consider $\alpha > 1$.
The sum in Eq.~\eqref{eq:transfer_runtime_bound_summation} is then $O(1)$ with respect to $r$.
Since $\gamma$ is arbitrary, Eq.~\eqref{eq:transfer_runtime_result} follows for any $z > 1$ (namely choose $\gamma < (z-1)/\alpha$).
Note that this conclusion also applies when $\mu(J)$ decays faster than a power law --- in such a case, $p_k$ (for $k \geq 2$) is even smaller than for any finite $\alpha$ and thus Eq.~\eqref{eq:transfer_runtime_bound_summation} remains a valid bound.

Next suppose $\alpha \leq 1$.
The sum now grows no faster than $O(r^{K(1 - \alpha) \gamma})$, and thus
\begin{equation} \label{eq:transfer_runtime_bound_tail}
T_r \leq Cr^{1 + \alpha \gamma + K (1 - \alpha) \gamma} \leq Cr^{\frac{1}{\alpha} + \gamma}.
\end{equation}
The latter inequality follows because, by definition, $1/\alpha \gamma < K \leq 1 + 1/\alpha \gamma$.
Again, since $\gamma$ is arbitrary, Eq.~\eqref{eq:transfer_runtime_result} follows.

Incidentally, this line of reasoning puts our discussion regarding the failure of the conventional LR bound (Eq.~\eqref{eq:sum_log_constraints_behavior} in particular) on firmer ground.
As noted above, the couplings enter into the conventional bound via the sum $\sum_{l=1}^r \log{J_l}$.
We have that
\begin{equation} \label{eq:sum_log_careful_bounds_v1}
r \sum_{k=0}^{\infty} p_k \log{a_{k+1}^{-1}} \leq \sum_{l=1}^r \log{J_l} \leq r \sum_{k=0}^{\infty} p_k \log{a_k^{-1}}.
\end{equation}
The sums over $k$ again terminate at $K$, but now the summands go as $r^{-k \alpha \gamma} \log{r^{-k \gamma}}$ and are dominated by small $k$ regardless of $\alpha$.
More precisely, using Eq.~\eqref{eq:interval_fraction_summary} and the analogous lower bound on $p_k$ gives
\begin{equation} \label{eq:sum_log_careful_bounds_v2}
Cr \log{r^{-\gamma}} \leq \sum_{l=1}^r \log{J_l} \leq C'r.
\end{equation}
Since $\gamma$ can be arbitrarily small, inserting into Eq.~\eqref{eq:standard_LR_bound} gives an LR bound whose front, while not necessarily quite ballistic, cannot have a dynamical exponent larger than 1.
As we have now established, that bound is far from tight.

\section{Disordered Lieb-Robinson bounds} \label{sec:disordered_bound}

As a non-trivial example of a situation in which the above results apply, here we suppose that each $J_l$ is drawn i.i.d.\ from a probability distribution $\mu(J)$ whose small-$J$ behavior is given by Eq.~\eqref{eq:empirical_distribution_tail_behavior}.
We first prove Eqs.~\eqref{eq:empirical_distribution_convergence_1} and~\eqref{eq:empirical_distribution_convergence_2}, not merely in some average sense but with probability 1, using standard techniques.
The results in Secs.~\ref{subsec:explicit_distribution_results} and~\ref{subsec:bound_tightness} then follow.

We next consider the threshold case $z = \max[1/\alpha, 1]$ in more detail.
For $\alpha > 1$, it follows immediately from the strong law of large numbers (see Refs.~\cite{Rosenthal2006,Durrett2019} for an introduction) that the transfer protocol in Sec.~\ref{subsec:bound_tightness} reaches all sites ballistically.
For $\alpha < 1$, on the other hand, the distinction between a.a.\ and i.o.\ bounds becomes important --- we derive an i.o.\ bound with arbitrarily small generalized velocity, implying that no protocol can reach every site in time $T_r = O(r^{1/\alpha})$, but also show that the above transfer protocol does reach an infinite \textit{subsequence} of sites in time $T_r = O(r^{1/\alpha})$ (again with probability 1).
Interestingly, $\alpha = 1$ is the only point at which we cannot give a definite answer.
Our i.o.\ bound still applies, but the transfer protocol now fails to reach \textit{any} site in time $T_r = O(r)$.
It could be that an a.a.\ bound with arbitrarily small velocity holds for $\alpha = 1$, but we have not succeeded in proving so.

Finally, we discuss some straightforward extensions of the above results.

\subsection{Convergence of the distribution} \label{subsec:disordered_distribution_convergence}

We first prove Eq.~\eqref{eq:empirical_distribution_convergence_1}.
This requires some tools from probability theory which can be found in textbooks on the subject~\cite{Rosenthal2006,Durrett2019} but are likely not common knowledge among physicists.
Here we apply these tools without further comment for ease of presentation, but include a description of them in App.~\ref{app:probabilistic_tools} for completeness.

\begin{widetext}
To prove Eq.~\eqref{eq:empirical_distribution_convergence_1}, pick any $\epsilon > 0$ and $\beta \in [0, 1/\alpha)$, then define the event $E_r$ to be
\begin{equation} \label{eq:empirical_distribution_convergence_1_events}
E_r \equiv \Big\{ \frac{\mu_r(J r^{-\beta})}{\mu(J r^{-\beta})} \not \in \big( 1 - \epsilon, 1 + \epsilon \big) \Big\}.
\end{equation}
Abbreviating $\mu(J r^{-\beta})$ by $\mu$ for conciseness, Eq.~\eqref{eq:empirical_distribution_convergence_1_events} is equivalently the event that the number of couplings less than $J r^{-\beta}$ is not between $(1 - \epsilon) r\mu$ and $(1 + \epsilon) r\mu$.
We can evaluate the latter directly:
\begin{equation} \label{eq:number_low_constraints_probability}
\begin{aligned}
\textrm{Pr} \big[ E_r \big] &= \sum_{n \not \in ((1 - \epsilon) r\mu, (1 + \epsilon) r\mu)} \binom{r}{n} \mu^n (1 - \mu)^{r - n} \\
&\leq \sum_{n \not \in ((1 - \epsilon) r\mu, (1 + \epsilon) r\mu)} \exp{\left[ n \log{\frac{r\mu}{n}} + (r - n) \log{\frac{r(1 - \mu)}{r - n}} \right]} \\
&\leq r \exp{\left[ -(1 + \epsilon) r\mu \log{(1 + \epsilon)} - \left( 1 - \frac{\mu}{1 - \mu} \epsilon \right) (1 - \mu) r \log{\left( 1 - \frac{\mu}{1 - \mu} \epsilon \right)} \right]},
\end{aligned}
\end{equation}
where the final inequality follows because the summand is maximized at $n = (1 + \epsilon) r\mu$.
One can confirm that the right-hand side goes as $r \exp{[-C r^{1 - \beta \alpha}]}$ at large $r$, with $C$ positive.
Since $1 - \beta \alpha > 0$, $\textrm{Pr}[E_r]$ is therefore summable and the probability of $E_r$ occurring infinitely often is zero (see App.~\ref{app:probabilistic_tools}).
The probability that this occurs for any rational $\epsilon$ or $\beta$, i.e., that $\mu_r(Jr^{-\beta}) / \mu(Jr^{-\beta})$ does not converge to 1 for any $\beta < 1/\alpha$, is likewise zero.

We now prove Eq.~\eqref{eq:empirical_distribution_convergence_2}.
Pick any $M > 0$ and $\beta > 1/\alpha$, and consider the event 
\begin{equation} \label{eq:extra_small_constraint_event}
E_r \equiv \Big\{ \min_{l=1}^r J_l \leq M r^{-\beta} \Big\}.
\end{equation}
We have (see App.~\ref{app:probabilistic_tools} for the definition of ``$E_r$ i.o.'')
\begin{equation} \label{eq:extra_small_constraint_probability_limit}
\textrm{Pr} \big[ E_r \textrm{ i.o.} \big] = \lim_{R \rightarrow \infty} \textrm{Pr} \left[ \exists r > R: \min_{l=1}^r J_l \leq M r^{-\beta} \right] = 1 - \lim_{R \rightarrow \infty} \textrm{Pr} \left[ \forall r > R: \min_{l=1}^r J_l > M r^{-\beta} \right].
\end{equation}
Clearly if $J_l > M r^{-\beta}$, then $J_l > M s^{-\beta}$ for all $s > r$.
Thus the following events are equivalent:
\begin{equation} \label{eq:constraint_bounds_equivalence}
\begin{aligned}
\Big\{ \forall r > R: \min_{l=1}^r J_l > Mr^{-\beta} \Big\} &= \Big\{ J_1 > M (R+1)^{-\beta} \Big\} \bigcap \Big\{ J_2 > M (R+1)^{-\beta} \Big\} \bigcap \cdots \bigcap \Big\{ J_{R+1} > M (R+1)^{-\beta} \Big\} \\
&\qquad \qquad \bigcap \Big\{ J_{R+2} > M (R+2)^{-\beta} \Big\} \bigcap \Big\{ J_{R+3} > M (R+3)^{-\beta} \Big\} \bigcap \cdots.
\end{aligned}
\end{equation}
Since the couplings $\{J_l\}$ are independent, the probability of the right-hand side is straightforward to evaluate:
\begin{equation} \label{eq:extra_small_constraint_evaluation}
\begin{aligned}
\textrm{Pr} \left[ \forall r > R: \min_{l=1}^r J_l > M r^{-\beta} \right] &= \Big( 1 - \mu \big( M(R+1)^{-\beta} \big) \Big)^{R+1} \lim_{R' \rightarrow \infty} \prod_{r=R+2}^{R'} \Big( 1 - \mu \big( Mr^{-\beta} \big) \Big) \\
&\sim \Big( 1 - C (R+1)^{-\beta \alpha} \Big)^{R+1} \exp{\left[ \sum_{r=R+2}^{\infty} \log{\Big( 1 - C r^{-\beta \alpha} \Big)} \right]},
\end{aligned}
\end{equation}
where the sum in the lower line is convergent because $\beta \alpha > 1$.
Therefore
\begin{equation} \label{eq:extra_small_constraint_result}
\lim_{R \rightarrow \infty} \textrm{Pr} \left[ \forall r > R: \min_{l=1}^r J_l > M r^{-\beta} \right] = 1,
\end{equation}
\end{widetext}
and $\textrm{Pr}[\min_{l=1}^r J_l \leq M r^{-\beta} \textrm{ i.o.}] = 0$.
The probability for any $M \in \mathbb{N}$ or rational $\beta$, i.e., the probability that $\min_{l=1}^r J_l / r^{-\beta}$ does not diverge, is likewise zero. This completes the proof of Eq.~\eqref{eq:empirical_distribution_convergence_2}.

\subsection{Fluctuations at the threshold exponent} \label{subsec:disordered_threshold_fluctuations}

We now take the dynamical exponent $z$ to be the threshold value $\max[1/\alpha, 1]$, still for the model in which all couplings $\{J_l\}$ are chosen i.i.d.\ from $\mu(J)$.
We determine whether an LR bound with arbitrarily small $v$ holds, both in the a.a.\ and i.o.\ sense.
In the situations where we can provide a decisive answer (which will be all $\alpha \neq 1$), this completes the diagram in Fig.~\ref{fig:main_result}.

First suppose $\alpha > 1$.
Then $\int \textrm{d}\mu(J) J^{-1}$ is finite (note that $\textrm{d}\mu(J) \sim C J^{\alpha - 1} \textrm{d}J$ at small $J$), and recall that we identified a specific protocol for which the runtime is $T_r = C \sum_{l=1}^r 1/J_l$.
Since the couplings are i.i.d., the strong law of large numbers (SLLN)~\cite{Rosenthal2006,Durrett2019} gives that with probability 1,
\begin{equation} \label{eq:large_alpha_runtime_convergence}
\lim_{r \rightarrow \infty} \frac{T_r}{r} = \int \textrm{d}\mu(J) J^{-1} < \infty.
\end{equation}
Thus the above transfer protocol reaches \textit{every} sufficiently large-distance site with a non-zero velocity (namely $[\int \textrm{d}\mu(J) J^{-1}]^{-1}$).

Now suppose $\alpha < 1$.
Return to Eq.~\eqref{eq:general_improved_LR_bound}, and take the subset $\lambda$ to be solely the link connecting $r-1$ and  $r$:
\begin{equation} \label{eq:threshold_improved_LR_bound}
\big\lVert \mathcal{P}_r A_0^t \big\rVert \leq 8 \big\lVert A_0 \big\rVert J_r t.
\end{equation}
Although a rather loose bound, the probability that even the right-hand side of Eq.~\eqref{eq:threshold_improved_LR_bound} exceeds $vtr^{-1/\alpha}$ for \textit{all} large $r$ vanishes, for any $v > 0$.
In other words, we have an i.o.\ bound with arbitrarily small generalized velocity: with probability 1, there exists a subsequence $\{r_i\}$ for which
\begin{equation} \label{eq:threshold_io_bound}
\big\lVert \mathcal{P}_{r_i} A_0^t \big\rVert \leq \big\lVert A_0 \big\rVert \frac{vt}{r_i^{1/\alpha}}.
\end{equation}
To prove Eq.~\eqref{eq:threshold_io_bound}, simply compute the probability that for all $r > R$, $J_r > vr^{-1/\alpha}$:
\begin{equation} \label{eq:threshold_io_evaluation}
\begin{aligned}
&\textrm{Pr} \Big[ \forall r > R: J_r > vr^{-1/\alpha} \Big] \\
&\qquad \qquad \qquad = \lim_{R' \rightarrow \infty} \prod_{r=R+1}^{R'} \Big( 1 - \mu \big( vr^{-1/\alpha} \big) \Big) \\
&\qquad \qquad \qquad \sim \lim_{R' \rightarrow \infty} \prod_{r=R+1}^{R'} \Big( 1 - C r^{-1} \Big) = 0.
\end{aligned}
\end{equation}

This establishes an i.o.\ bound with arbitrarily small $v$, but let us consider again the previous transfer protocol.
One can show that with probability 1,
\begin{equation} \label{eq:threshold_transfer_protocol_runtime}
\liminf_{r \rightarrow \infty} \frac{T_r}{r^{1/\alpha}} = 0,
\end{equation}
meaning that this protocol does reach a \textit{subsequence} of sites $\{r_i\}$ in time of order (and in fact asymptotically smaller than) $r_i^{1/\alpha}$.
The proof of Eq.~\eqref{eq:threshold_transfer_protocol_runtime}, which we have adapted from Ref.~\cite{liminf_proof}, begins by noting that $T_r r^{-1/\alpha}$ converges in distribution to a non-negative random variable $S$ whose support includes 0 (a fact which is well-established but by no means trivial, e.g., see Ref.~\cite{Durrett2019}).
Thus pick any $\epsilon > 0$.
We have that $\textrm{Pr}[S \geq \epsilon] < 1$, and at the same time, we can always choose a sequence of numbers $\{C_k\}$ such that $\textrm{Pr}[S \geq C_k] < 1/k^2 \rightarrow 0$.

The convergence in distribution of $T_r r^{-1/\alpha}$ to $S$ implies that we can construct a subsequence $\{r_k\}$ with the following properties:
\begin{itemize}
\item $r_{k+1} - r_k \rightarrow \infty$ as $k \rightarrow \infty$;
\item For sufficiently large $k$,
\begin{equation} \label{eq:threshold_runtime_convergence_property}
\bigg| \textrm{Pr} \Big[ T_{r_k} r_k^{-1/\alpha} \geq C_k \Big] - \textrm{Pr} \Big[ S \geq C_k \Big] \bigg| < \frac{1}{k^2};
\end{equation}
\item $\epsilon_k > \epsilon$, where we define (for later convenience)
\begin{equation} \label{eq:threshold_runtime_convergence_epsilon}
\epsilon_k \equiv \frac{2 \epsilon r_{k+1}^{1/\alpha} - C_k r_k^{1/\alpha}}{(r_{k+1} - r_k)^{1/\alpha}}.
\end{equation}
\end{itemize}
From Eq.~\eqref{eq:threshold_runtime_convergence_property}, it follows that $\textrm{Pr}[T_{r_k} r_k^{-1/\alpha} \geq C_k] < 2/k^2$, and thus with probability 1, there is some $K$ such that $T_{r_k} r_k^{-1/\alpha} < C_k$ for all $k > K$.

Consider, for $k > K$, the event
\begin{equation} \label{eq:threshold_runtime_slow_event}
E_k \equiv \Big\{ \forall k' \geq k: T_{r_{k'}} \geq 2 \epsilon r_{k'}^{1/\alpha} \Big\}.
\end{equation}
Since $T_{r_{k'}} < C_{k'} r_{k'}^{1/\alpha}$ with probability 1, we have that
\begin{equation} \label{eq:threshold_runtime_slow_bound_derivation}
\begin{aligned}
&\textrm{Pr} \left[ \forall k' \geq k: T_{r_{k'}} \geq 2 \epsilon r_{k'}^{1/\alpha} \right] \\
&\; = \textrm{Pr} \left[ \forall k' \geq k: T_{r_{k'}} \geq 2 \epsilon r_{k'}^{1/\alpha} \bigcap T_{r_{k'}} < C_{k'} r_{k'}^{1/\alpha} \right] \\
&\; \leq \textrm{Pr} \left[ \forall k' \geq k: T_{r_{k'+1}} - T_{r_{k'}} > 2 \epsilon r_{k'+1}^{1/\alpha} - C_{k'} r_{k'}^{1/\alpha} \right] \\
&\; = \textrm{Pr} \left[ \forall k' \geq k: \frac{T_{r_{k'+1}} - T_{r_{k'}}}{(r_{k'+1} - r_{k'})^{1/\alpha}} > \epsilon_{k'} \right] \\
&\; \leq \textrm{Pr} \left[ \forall k' \geq k: \frac{T_{r_{k'+1}} - T_{r_{k'}}}{(r_{k'+1} - r_{k'})^{1/\alpha}} > \epsilon \right],
\end{aligned}
\end{equation}
using Eq.~\eqref{eq:threshold_runtime_convergence_epsilon}.
Note that the \textit{differences} $T_{r_{k'+1}} - T_{r_{k'}}$ are mutually independent, and thus the probability on the right-hand side factors:
\begin{equation} \label{eq:threshold_runtime_slow_bound}
\textrm{Pr} \big[ E_k \big] \leq \lim_{K' \rightarrow \infty} \prod_{k'=k}^{K'} \textrm{Pr} \left[ \frac{T_{r_{k'+1}} - T_{r_{k'}}}{(r_{k'+1} - r_{k'})^{1/\alpha}} > \epsilon \right].
\end{equation}
Furthermore, since $r_{k+1} - r_k \rightarrow \infty$, the random variable $(T_{r_{k+1}} - T_{r_k})/(r_{k+1} - r_k)^{1/\alpha}$ itself converges in distribution to $S$, and so for sufficiently large $k$,
\begin{equation} \label{eq:threshold_runtime_increment_bound}
\textrm{Pr} \left[ \frac{T_{r_{k+1}} - T_{r_k}}{(r_{k+1} - r_k)^{1/\alpha}} > \epsilon \right] < \frac{1 + \textrm{Pr}[S \geq \epsilon]}{2}.
\end{equation}
The right-hand side is strictly less than one, meaning the infinite product in Eq.~\eqref{eq:threshold_runtime_slow_bound} evaluates to zero.
Thus $\textrm{Pr}[E_k] = 0$ and therefore
\begin{equation} \label{eq:threshold_runtime_final_result}
\textrm{Pr} \left[ T_{r_k} < 2 \epsilon r_k^{1/\alpha} \textrm{ i.o.} \right] = 1 - \lim_{k \rightarrow \infty} \textrm{Pr} \big[ E_k \big] = 1.
\end{equation}
In other words, Eq.~\eqref{eq:threshold_transfer_protocol_runtime} holds with probability 1.

It remains only to consider $\alpha = 1$.
The calculation in Eq.~\eqref{eq:threshold_io_evaluation} still holds, and thus an i.o.\ bound with arbitrarily small velocity exists with probability 1.
Yet for the transfer protocol, we now have
\begin{equation} \label{eq:marginal_case_transfer_runtime_divergence}
\liminf_{r \rightarrow \infty} \frac{T_r}{r} = \infty,
\end{equation}
i.e., \textit{no} site is reached ballistically.
Compare to Eq.~\eqref{eq:threshold_transfer_protocol_runtime}.
It may be that a more sophisticated protocol is able to reach a subsequence ballistically, or it may be that no such protocol exists.
We have been unable to rule out either possibility.

To prove Eq.~\eqref{eq:marginal_case_transfer_runtime_divergence}, following Ref.~\cite{Durrett2019}, pick any $M > 0$ and define truncated random variables $Y_l' \equiv \min[Y_l, M]$ (recall that $Y_l \equiv 1/J_l$).
The expectation value of $Y_l'$, denoted $\overline{Y}(M)$, is finite and therefore the SLLN applies.
Thus with probability 1,
\begin{equation} \label{eq:marginal_case_truncated_convergence}
\liminf_{r \rightarrow \infty} \frac{T_r}{r} = \liminf_{r \rightarrow \infty} \frac{1}{r} \sum_{l=1}^r Y_l \geq \liminf_{r \rightarrow \infty} \frac{1}{r} \sum_{l=1}^r Y_l' = \overline{Y}(M).
\end{equation}
Since $\overline{Y}(M) \rightarrow \infty$ as $M \rightarrow \infty$, Eq.~\eqref{eq:marginal_case_transfer_runtime_divergence} follows.

\subsection{Extensions} \label{eq:subsec:disordered_extensions}

Lastly, we discuss some straightforward extensions of the above results. These are not meant to be exhaustive, nor do we expect them to be particularly tight bounds --- we only wish to point out some generalizations that can be obtained with little additional work.

\textbf{Couplings with finite-range correlations:}
As a first example, suppose that the couplings $\{J_l\}$ are correlated, but that correlations exist only within a finite range $\xi$.
By the latter, we mean that joint distributions $\mu^{(n)}$ factor only if all couplings involved are separated by at least $\xi$ sites, e.g.,
\begin{equation} \label{eq:correlated_constraints_factoring_example}
\begin{aligned}
&\mu^{(n+1)} \big( J_l, J_{l+\xi}, \cdots, J_{l+n\xi} \big) \\
&\qquad \qquad \qquad = \mu^{(1)}(J_l) \mu^{(1)}(J_{l+\xi}) \cdots \mu^{(1)}(J_{l+n\xi}).
\end{aligned}
\end{equation}
Such correlations present no difficulties --- simply first pass to the interaction picture with respect to all but every $\xi$'th link, and then the previous analysis applies.
All lengths are reduced by a factor of $\xi$, and thus the generalized LR velocity is increased by a factor of $\xi^z$ (which we again do not claim to be a particularly accurate estimate), but the diagram in Fig.~\ref{fig:main_result} remains unmodified.

\textbf{Bounds for multiple energy scales:}
Returning to a strictly 1D chain, suppose that $J_l$ can only take the values $J$ and $\epsilon J$, where $0 < \epsilon \ll 1$.
The couplings are still chosen independently, and the probability of $J_l = \epsilon J$ is $\epsilon^{\alpha}$.
This is a discrete analogue to the situation from the previous subsections, for which $J_l$ could take any value greater than zero.
Here, any LR bound will clearly have a ballistic front, but one can still ask how the LR velocity $v$ compares to the two scales $J$ and $\epsilon J$.
By taking $\lambda$ in Eq.~\eqref{eq:general_improved_LR_bound} to be those links with $J_l = \epsilon J$ (the fraction of which approaches $\epsilon^{\alpha}$ at large $r$ with probability 1), and comparing to the conventional LR bound, we find that
\begin{equation} \label{eq:two_scale_velocity_scaling}
v = 4e \epsilon^{\max[1 - \alpha, 0]} J.
\end{equation}
Analogous to the previous results, $v \ll J$ for $\alpha < 1$.
Furthermore, the dependence of $v$ on $\epsilon$ in Eq.~\eqref{eq:two_scale_velocity_scaling} is tight --- the average of $1/J_l$ is finite for all $\alpha$, namely given by $(1 - \epsilon^{\alpha} + \epsilon^{\alpha - 1})/J$, and so the SLLN again applies, as in Eq.~\eqref{eq:large_alpha_runtime_convergence}.
Taking $\epsilon \ll 1$, the velocity of the transfer protocol is therefore Eq.~\eqref{eq:two_scale_velocity_scaling} up to prefactors.

We can easily generalize to there being an arbitrary (finite) number of widely separated energy scales.
Suppose that $J_l = \epsilon^{\gamma_k} J$ with probability $\epsilon^{\alpha_k}$ for $k \in \{1, \cdots, K\}$, and $J_l = J$ otherwise.
Assume $0 < \gamma_1 < \cdots < \gamma_K$ and $0 < \alpha_1 < \cdots < \alpha_K$.
The optimal LR velocity is now
\begin{equation} \label{eq:many_scale_velocity_scaling}
\begin{gathered}
v = 4e \epsilon^{\zeta} J, \\
\zeta \equiv \max[\gamma_K - \alpha_K, \cdots, \gamma_1 - \alpha_1, 0].
\end{gathered}
\end{equation}
The dependence on $\epsilon$ is again tight.
While simple, this result does highlight that in general, neither the largest nor the smallest energy scale necessarily determines the relevant velocity for operator growth on its own.

\begin{figure}[t]
\centering
\includegraphics[width=1.0\columnwidth]{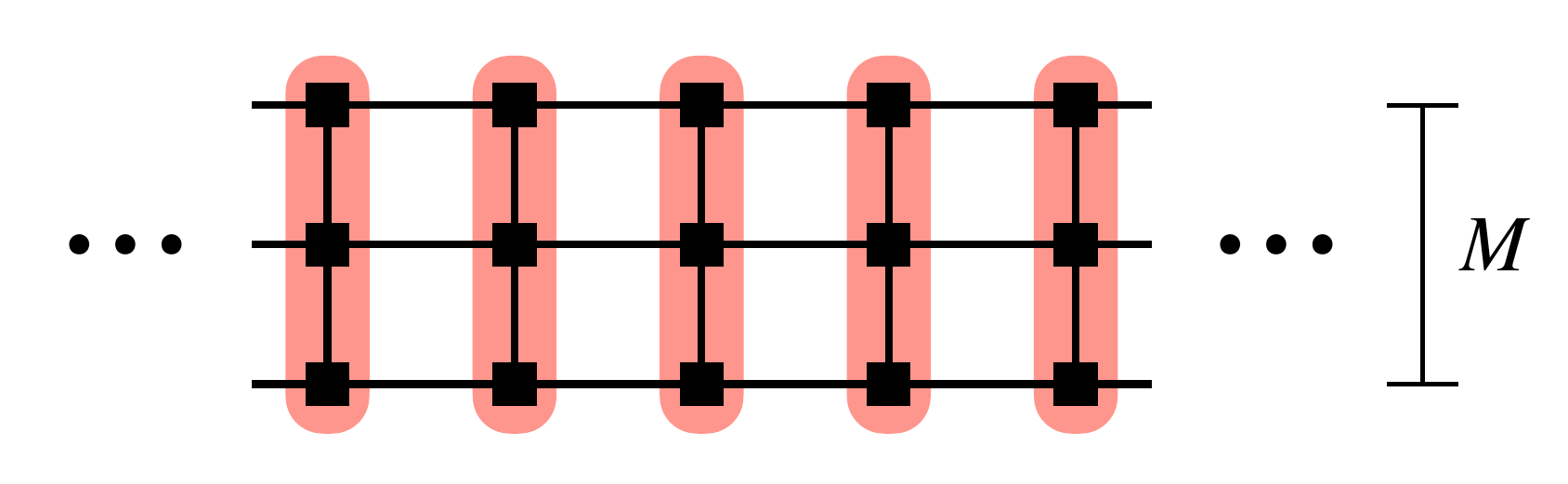}
\caption{Example of a ``ladder'', with $M = 3$ sites per rung. To apply our analysis, simply treat all $M$ sites connected vertically as a single ``site'', indicated by the red shading.}
\label{fig:1D_chain}
\end{figure}

\textbf{Bounds for ladders:} Consider a system such as in Fig.~\ref{fig:1D_chain}, in which sites are labelled by $(i,j)$ with $i \in \{ \cdots, -1, 0, 1, \cdots \}$ and $j \in \{1, \cdots , M\}$.
The Hamiltonian is still given by a sum of terms for each link of the lattice.
Interactions along vertical links are arbitrary, and interactions along horizontal links (denoted $H_{ij}(t)$ for the link between $(i-1,j)$ and $(i,j)$) obey $\lVert H_{ij}(t) \rVert \leq J_{ij}$.
Each $J_{ij}$ is again drawn independently from $\mu(J)$.

Since our analysis does not make any assumptions regarding the nature of the local Hilbert space, we can simply identify each set of sites connected vertically --- $\{(i,j)\}_{j=1}^M$ for fixed $i$ --- as comprising a single ``local'' Hilbert space.
However, the interaction between neighboring $i$ is then $\sum_j H_{ij}(t)$, meaning that the coefficient $J_l$ entering into bounds such as Eq.~\eqref{eq:general_improved_LR_bound} should be $\sum_j J_{lj}$.
The probability of $\sum_j J_{lj} \leq J$ is bounded by
\begin{equation} \label{eq:coefficient_sum_bounds}
\mu \big( J/M \big)^M \leq \textrm{Pr} \left[ \sum_j J_{lj} \leq J \right] \leq \mu \big( J \big)^M.
\end{equation}
Thus the correct exponent to use in our analysis is now $M \alpha$, and in particular, $z_c(\alpha) = \max [1/M \alpha, 1]$.
Obtaining an improvement over the conventional LR bound now requires $\alpha < 1/M$, but nonetheless, we still have that $z_c \rightarrow \infty$ as $\alpha \rightarrow 0$.

In fact, we can adapt the construction of Ref.~\cite{Tran2021Optimal} to show that this result for $z_c(\alpha)$ is tight.
We discuss this assuming two states per site ($d = 2$) labelled by $|0 \rangle$ and $|1 \rangle$ --- the same protocol extends to arbitrary $d$ simply by acting as the identity in the subspace orthogonal to $|0 \rangle$ and $|1 \rangle$, and it still produces an $O(1)$ commutator for generic operators $A_0$ and $B_r$ which act non-trivially on $|0 \rangle$ and $|1 \rangle$.

Taking cues from Ref.~\cite{Tran2021Optimal}, consider starting in the product state with $a |0 \rangle + b |1 \rangle$ on site $(i-1, j)$ and $|0 \rangle$ on all other sites of rungs $i-1$ and $i$.
Since arbitrary vertical interactions are allowed, we can construct a unitary that takes this state to $(a |\overline{0} \rangle_{i-1} + b |\overline{1} \rangle_{i-1}) \otimes (|\overline{0} \rangle_i + |\overline{1} \rangle_i)/\sqrt{2}$ in arbitrarily short time, where $|\overline{0} \rangle_i$ and $|\overline{1} \rangle_i$ denote the states on rung $i$ with all sites in $|0 \rangle$ and $|1 \rangle$ respectively.
Defining $H_i \equiv \sum_{j=1}^M J_{ij} |1 \rangle \langle 1|_{(i-1,j)} \otimes |1 \rangle \langle 1|_{(i,j)}$, we have
\begin{equation} \label{eq:ladder_state_transfer}
\begin{aligned}
&e^{i \pi H_i / \sum_j J_{ij}} \Big( a \big| \overline{0} \big>_{i-1} + b \big| \overline{1} \big>_{i-1} \Big) \otimes \frac{\big| \overline{0} \big>_i + \big| \overline{1} \big>_i}{\sqrt{2}} \\
&\; = a \big| \overline{0} \big>_{i-1} \otimes \frac{\big| \overline{0} \big>_i + \big| \overline{1} \big>_i}{\sqrt{2}} + b \big| \overline{1} \big>_{i-1} \otimes \frac{\big| \overline{0} \big>_i - \big| \overline{1} \big>_i}{\sqrt{2}}.
\end{aligned}
\end{equation}
The states $(|\overline{0} \rangle_i + |\overline{1} \rangle_i)/\sqrt{2}$ and $(|\overline{0} \rangle_i - |\overline{1} \rangle_i)/\sqrt{2}$ can then be converted into $|\overline{0} \rangle_i$ and $|\overline{1} \rangle_i$ respectively, again in arbitrarily short time, using interactions solely on rung $i$.
This procedure thus transforms the product state having $a |0 \rangle + b |1 \rangle$ on site $(i-1,j)$ and $|0 \rangle$ otherwise into the generalized GHZ state $a |\overline{0} \rangle_{i-1} \otimes |\overline{0} \rangle_i + b |\overline{1} \rangle_{i-1} \otimes |\overline{1} \rangle_i$.
Subsequently applying the procedure in reverse, albeit with the roles of rungs $i-1$ and $i$ exchanged, then takes this generalized GHZ state into the product state having $a |0 \rangle + b |1 \rangle$ on site $(i,j)$ and $|0 \rangle$ otherwise.
The net effect is that the state on site $(i-1, j)$ has been transferred to site $(i, j)$ in a time $2 \pi/\sum_j J_{ij}$ (coming from Eq.~\eqref{eq:ladder_state_transfer}).
Repeating the transfer sequentially from rung 0 to $r$, we have a protocol analogous to Fig.~\ref{fig:transfer_protocol} with runtime $T_r = \sum_{l=1}^r 2 \pi/\sum_j J_{ij}$.

Thus not only does our LR bound apply to the ladder of Fig.~\ref{fig:1D_chain}, with $\sum_j J_{lj}$ in place of $J_l$, but so does our analysis of the 1D transfer protocol, again using $\sum_j J_{lj}$ as an effective horizontal coupling.
The result derived above that $z_c(\alpha) = \max[1/M \alpha, 1]$ is therefore tight.
Note that if we restrict ourselves to bounded-strength vertical interactions, but with bounds that are spatially uniform, then the dynamical exponent remains unaffected even once the time required to effect all single-rung transformations is incorporated.
We leave the more complicated situation in which the vertical interactions themselves have weak links as a direction for future work.

The fact that $z_c(\alpha) \rightarrow 1$ as $M \rightarrow \infty$ for any $\alpha > 0$ suggests that our conclusions may not extrapolate to higher dimensions (and analogously to longer-range interactions).
There are far more paths connecting any two sites in higher dimensions, and it may be that transport remains ballistic for any power-law distribution of weak links.
Of course, the analysis of the ladder presented here only accounts for weak links in one direction, and so the behavior of truly multi-dimensional disordered systems remains an important open question.

\section{Applications} \label{sec:applications}

In Sec.~\ref{sec:general_bound}, we derived a modified LR bound for non-translation-invariant systems --- Eq.~\eqref{eq:general_tail_upper_bound_v1} --- requiring only that the empirical distribution $\mu_r(J)$ converge to a function $\mu(J) \sim CJ^{\alpha}$ (as formalized by Eqs.~\eqref{eq:empirical_distribution_convergence_1} and~\eqref{eq:empirical_distribution_convergence_2}).
For $\alpha < 1$, the modified bound gives a significant improvement over the conventional bound, and even guarantees that operator spreading is sub-ballistic.
Here we consider the consequences of this result for various applications (assume $\alpha < 1$ throughout).

On the one hand, the manner in which LR bounds are used in the following is quite similar from case to case.
Yet we shall see that different contexts come with different caveats, some more than others, and we discuss the many open directions for future work.

Note as well that we are working with the general bound of Sec.~\ref{sec:general_bound} rather than the more detailed results of Sec.~\ref{sec:disordered_bound}.
In particular, implications of the ``a.a.''-``i.o.'' distinction for the following applications warrant further investigation.

\subsection{Growth of correlations} \label{subsec:correlation_growth}

LR bounds directly place limitations on the extent to which correlations can develop following a quench.
Consider the correlation function
\begin{equation} \label{eq:correlation_function_definition}
G(t) \equiv \big< A^t B^t \big> - \big< A^t \big> \big< B^t \big>,
\end{equation}
where the expectation value is in a product state $|\Psi \rangle$, time evolution is under a Hamiltonian $H(t) \in \mathcal{H}_J$, and the operators $A^0$ and $B^0$ are supported on sites to the left of 0 and to the right of $r$ respectively (note that our analysis in Sec.~\ref{sec:general_bound} applies equally well to such operators even if they are not strictly local).
Thus $G(0) = 0$, and one would like to understand how $G(t)$ grows in time.

The authors of Ref.~\cite{Bravyi2006LiebRobinson} show that one can bound (assuming $\lVert A \rVert = \lVert B \rVert = 1$ for simplicity)
\begin{equation} \label{eq:correlation_function_general_bound}
\big| G(t) \big| \leq 4 \big\lVert \mathcal{P}_{\geq r/2} A^t \big\rVert + 4 \big\lVert \mathcal{P}_{\leq r/2} B^t \big\rVert.
\end{equation}
Supposing that $r$ is sufficiently large, we can immediately apply Eq.~\eqref{eq:general_tail_upper_bound_v1}:
\begin{equation} \label{eq:correlation_function_bound_v1}
\big| G(t) \big| \leq \exp{\left[ -Cr^{1 - \beta \alpha} + \frac{C't}{r^{\beta}} \right]},
\end{equation}
where $\beta$ can take any value in $(0, 1/\alpha)$.
The fact that one can choose an optimal $\beta$ depending on $t$ makes Eq.~\eqref{eq:correlation_function_bound_v1} slightly more interesting than the conventional bound.

\begin{figure}[t]
\centering
\includegraphics[width=1.0\columnwidth]{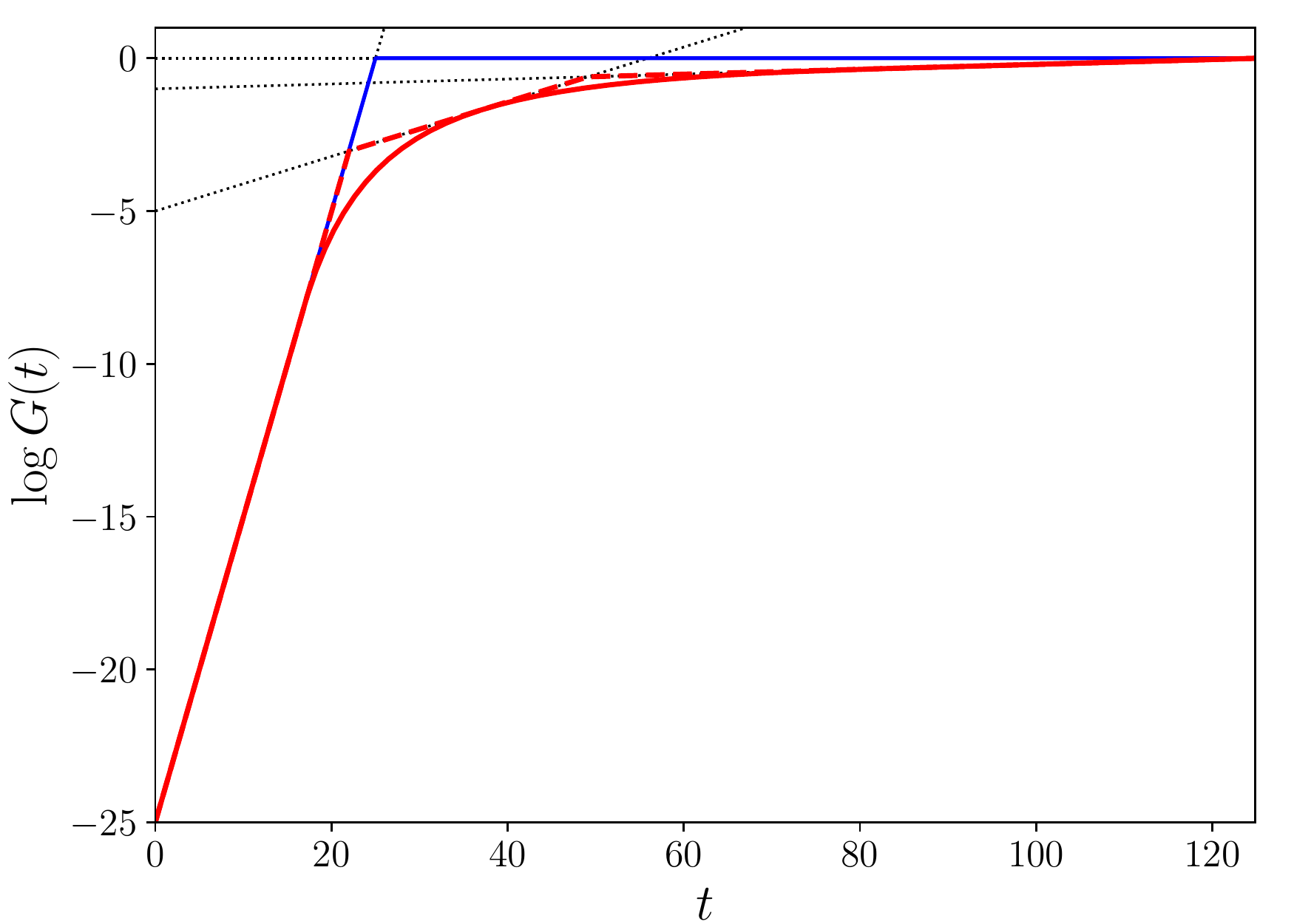}
\caption{Plot of Eq.~\eqref{eq:correlation_function_bound_v2}, using $\alpha = 2/3$ for concreteness. Dashed red line corresponds to $n = 2$ with $\beta_i \in \{0, 3/4, 3/2\}$ (each of the individual curves being minimized among in Eq.~\eqref{eq:correlation_function_bound_v2} is shown as a dotted black line). Solid red line corresponds to $n = 30$ and $\beta_i$ equally spaced by $1/20$, which closely approximates the curve at large $n$. Blue line is the conventional LR bound for comparison. To reproduce the exact numbers in the plot, set $r = 25$ and $C = C' = 1$.}
\label{fig:correlator_growth_bound}
\end{figure}

However, care must be taken in varying $\beta$.
We have shown that for any $\beta \in (0, 1/\alpha)$, there exists $R(\beta)$ such that the bound holds for $r > R(\beta)$, but we have not shown that the convergence is uniform in $\beta$ (i.e., that $R(\beta)$ can be made independent of $\beta$).
Rather than impose an additional requirement on the convergence of $\mu_r(J)$ and attempt to verify it in non-trivial situations, here we shall simply choose a finite set $\{\beta_i\}_{i=0}^n$ with $0 \equiv \beta_0 < \cdots < \beta_n < 1/\alpha$ (note that $\beta = 0$ simply recovers the conventional LR bound).
The precise statement of Eq.~\eqref{eq:correlation_function_bound_v1} is that for $r/2 > \max[R(\beta_0), \cdots, R(\beta_n)]$,
\begin{equation} \label{eq:correlation_function_bound_v2}
\big| G(t) \big| \leq \exp{\Bigg[ \min_{i=0}^n \left[ -Cr^{1 - \beta_i \alpha} + \frac{C't}{r^{\beta_i}} \right] \Bigg]}.
\end{equation}
The behavior of Eq.~\eqref{eq:correlation_function_bound_v2} is shown in the dashed red line of Fig.~\ref{fig:correlator_growth_bound}.
At large $r$, it is given by $\exp{[-Cr + C't]}$ until $t = O(r)$, then by $\exp{[-Cr^{1 - \beta_1 \alpha} + C't/r^{\beta_1}]}$ until $t = O(r^{1 + \beta_1 (1 - \alpha)})$, and in general,
\begin{equation} \label{eq:correlation_function_bound_approximate}
\begin{aligned}
\big| G(t) \big| &\leq \exp{\left[ -Cr^{1 - \beta_i \alpha} + \frac{C't}{r^{\beta_i}} \right]}, \\
&\qquad \qquad \qquad \quad r^{1 + \beta_{i-1} (1 - \alpha)} \ll t \ll r^{1 + \beta_i (1 - \alpha)}.
\end{aligned}
\end{equation}
Of course, we are free to choose as large a set $\{\beta_i\}$ as we like (although this may increase the distance required for the bound to hold).
One should heuristically think of Eq.~\eqref{eq:correlation_function_bound_v2} as minimizing over all $\beta \in (0, 1/\alpha)$ for any value of $r$, shown as the solid red line in Fig.~\ref{fig:correlator_growth_bound}.

\subsection{Creation of topological order} \label{subsec:topological_order}

A related application is lower bounds on the time needed to create topological order.
Again following Ref.~\cite{Bravyi2006LiebRobinson}, we say that two states $|\Psi_1 \rangle$ and $|\Psi_2 \rangle$ in a 1D system of size $N$ are ``topologically ordered'' (relative to each other) if there exist constants $c_1 \in (0, 1)$ and $c_2 > 0$ such that, for every observable $O$ supported on a set with diameter $c_1 N$ or less,
\begin{equation} \label{eq:topological_order_definition}
\begin{gathered}
\Big| \big< \Psi_1 \big| O \big| \Psi_1 \big> - \big< \Psi_2 \big| O \big| \Psi_2 \big> \Big| < 2 e^{-c_2 N}, \\
\Big| \big< \Psi_1 \big| O \big| \Psi_2 \big> \Big| < e^{-c_2 N}.
\end{gathered}
\end{equation}
In words, no ``local'' operator (even one supported on a non-vanishing fraction of the system) can distinguish between or couple such states.

Suppose that $|\Psi_1 \rangle$ and $|\Psi_2 \rangle$ are topologically ordered and have been prepared from states $|\Phi_1 \rangle$ and $|\Phi_2 \rangle$ via time evolution under a Hamiltonian $H(t) \in \mathcal{H}_J$:
\begin{equation} \label{eq:topological_state_preparation_procedure}
\big| \Psi_i \big> = \mathcal{T} e^{-i \int_0^t \textrm{d}s H(s)} \big| \Phi_i \big>.
\end{equation}
Ref.~\cite{Bravyi2006LiebRobinson} shows that, for any operator $O$ with diameter less than $c_1 N/2$,
\begin{equation} \label{eq:topological_order_preparation_relation}
\Big| \big< \Phi_1 \big| O \big| \Phi_1 \big> - \big< \Phi_2 \big| O \big| \Phi_2 \big> \Big| \leq 2e^{-c_2 N} + 2 \big\lVert \mathcal{P}_{\geq c_1 N} O^t \big\rVert,
\end{equation}
and analogously for the off-diagonal matrix elements.
We again use that
\begin{equation} \label{eq:topological_spreading_bound}
\big\lVert \mathcal{P}_{\geq c_1 N} O^t \big\rVert \leq \exp{\Bigg[ \min_{i=0}^n \left[ -CN^{1 - \beta_i \alpha} + \frac{C't}{N^{\beta_i}} \right] \Bigg]}.
\end{equation}
Note that for $t \ll N$, Eq.~\eqref{eq:topological_spreading_bound} is exponentially small in $N$.
Returning to Eq.~\eqref{eq:topological_order_preparation_relation}, $|\Phi_1 \rangle$ and $|\Phi_2 \rangle$ then satisfy the definition of topological order.
In other words, for times less than $O(N)$, it is impossible to prepare topologically ordered states ($|\Psi_1 \rangle$ and $|\Psi_2 \rangle$) from any states which are not themselves topologically ordered ($|\Phi_1 \rangle$ and $|\Phi_2 \rangle$).

The above conclusion is identical to that of the conventional case, but note that Eq.~\eqref{eq:topological_spreading_bound} in fact remains small for much longer times, until $t = O(N^{1/\alpha})$.
The bound is no longer exponential (rather stretched-exponential), and so the definition in Eq.~\eqref{eq:topological_order_definition} is not strictly met, but $|\Phi_1 \rangle$ and $|\Phi_2 \rangle$ exhibit a slightly looser sense of topological order nonetheless.
In this sense, we have that for non-translation-invariant systems with $\alpha < 1$, it is impossible to prepare ordered states from unordered in times less than $O(N^{1/\alpha})$.

\subsection{Heating in periodically driven systems} \label{subsec:heating}

The authors of Ref.~\cite{Abanin2015Exponentially} consider energy absorption in weakly driven systems, for which the Hamiltonian is of the form
\begin{equation} \label{eq:energy_absorption_Hamiltonian_form}
H(t) = \sum_{l \in \Lambda} H_l - g \cos{\omega t} \sum_i O_i.
\end{equation}
We shall limit ourselves to 1D, although Ref.~\cite{Abanin2015Exponentially} treats a general dimension.
Within linear response theory~\cite{Chaikin1995}, i.e., to leading order at small $g$, the energy absorption rate is proportional to $\sigma(\omega) \equiv \sum_{ij} \sigma_{ij}(\omega)$, where
\begin{equation} \label{eq:local_response_function_definition}
\sigma_{ij}(\omega) = \frac{1}{2} \int_{-\infty}^{\infty} \textrm{d}t e^{i \omega t} \Big< \big[ O_i^t, O_j^0 \big] \Big>.
\end{equation}
The expectation value denoted by $\langle \, \cdot \, \rangle$ is taken in the initial (potentially mixed) state of the system.
Strictly speaking, $\sigma_{ij}(\omega)$ is a distribution and should be integrated against test functions to obtain meaningful results.

The authors first derive a bound
\begin{equation} \label{eq:energy_absorption_initial_bound}
\big| \sigma_{ij}(\omega) \big| \leq C e^{-\kappa |\omega|},
\end{equation}
assuming a translation-invariant constraint $\lVert H_l \rVert \leq J$.
There is already the potential for tightening Eq.~\eqref{eq:energy_absorption_initial_bound} in non-translation-invariant systems, since the constants $C$ and $\kappa$ involve sums over connected paths with factors of $J_l$, much as in Sec.~\ref{subsec:review_old_bound}.
However, we suspect that a more careful analysis of the path sum would not yield a significant improvement (see Eq.~\eqref{eq:sum_log_constraints_behavior}), and so we continue to use Eq.~\eqref{eq:energy_absorption_initial_bound}.

LR bounds enter into the analysis of Ref.~\cite{Abanin2015Exponentially} as a means of bounding $|\sigma(\omega)|$ by $O(N)$ ($N$ being the number of sites in the system --- note that naively bounding $|\sigma(\omega)| \leq \sum_{ij} |\sigma_{ij}(\omega)|$ would give a bound $O(N^2)$).
In particular, consider two sites $i$ and $j$ separated by a distance $r > r^*$, with $r^*$ to be chosen later.
Assuming a conventional LR bound with velocity $v$, Eq.~\eqref{eq:local_response_function_definition} can be bounded by two contributions:
\begin{equation} \label{eq:distant_response_function_bound}
\begin{aligned}
\big| \sigma_{ij}(\omega) \big| &\leq C \int_0^{r/v} \textrm{d}t e^{-\frac{t^2}{\delta t^2}} e^{-a(r - vt)} + C' \int_{r/v}^{\infty} \textrm{d}t e^{-\frac{t^2}{\delta t^2}} \\
&\leq C e^{-ar} + C' e^{-\frac{r^2}{v^2 \delta t^2}}.
\end{aligned}
\end{equation}
See Ref.~\cite{Abanin2015Exponentially} for details, including the appearance of the Gaussian factor.
Since the terms of Eq.~\eqref{eq:distant_response_function_bound} decay exponentially with $r$ or faster, the sum over all $i$ and $j$ with $|i - j| > r^*$ is indeed $O(N)$ and scales as $\exp{[-ar^*]}$.
For summing over $|i - j| < r^*$, simply use Eq.~\eqref{eq:energy_absorption_initial_bound}.
Thus by setting $r^* = O(|\omega|)$, both contributions decay exponentially with $|\omega|$, and therefore
\begin{equation} \label{eq:energy_absorption_final_bound}
\big| \sigma(\omega) \big| \leq CN e^{-\kappa' |\omega|}.
\end{equation}
The exponential decay with $|\omega|$ is the main result of Ref.~\cite{Abanin2015Exponentially}.

Let us consider whether this conclusion is altered in non-translation-invariant systems by the use of our modified LR bound.
As in the preceding subsections, we choose a set $\{ \beta_i \}_{i=0}^n$ and minimize Eq.~\eqref{eq:general_tail_upper_bound_v1} over $\beta_i$.
The integral over $t$ in Eq.~\eqref{eq:local_response_function_definition} splits into multiple terms (compare to Eq.~\eqref{eq:distant_response_function_bound}):
\begin{equation} \label{eq:modified_distant_response_function_bound}
\begin{aligned}
\big| \sigma_{ij}(\omega) \big| &\leq C_0 \int_0^{r/v_0} \textrm{d}t e^{-\frac{t^2}{\delta t^2}} e^{-a_0 (r - v_0 t)} \\
&+ C_1 \int_{r/v_0}^{r^{1 + \beta_1 (1 - \alpha)}/v_1} \textrm{d}t e^{-\frac{t^2}{\delta t^2}} e^{-a_1 (r^{1 - \beta_1 \alpha} - v_1 t r^{-\beta_1})} \\
&+ C_2 \int_{r^{1 + \beta_1 (1 - \alpha)}/v_1}^{r^{1 + \beta_2 (1 - \alpha)}/v_2} \textrm{d}t e^{-\frac{t^2}{\delta t^2}} e^{-a_2 (r^{1 - \beta_2 \alpha} - v_2 t r^{-\beta_2})} \\
&+ \cdots + C_n \int_{r^{1 + \beta_n (1 - \alpha)}/v_n}^{\infty} \textrm{d}t e^{-\frac{t^2}{\delta t^2}}.
\end{aligned}
\end{equation}
Although the latter terms are indeed much smaller than in the translation-invariant case, note that the first term is unaffected.
Thus $|\sigma_{ij}(\omega)|$ still scales as $\exp{[-ar]}$ at large $r$, we are still led to take $r^* = O(|\omega|)$, and the final result in Eq.~\eqref{eq:energy_absorption_final_bound} is unchanged.

The lack of any significant reduction in the heating rate can be traced back to the fact that it is the \textit{tail} of the LR bound which constrains $|\sigma(\omega)|$, and the tail of the non-translation-invariant LR bound is no tighter than that of the translation-invariant case.
Of course, this is only a statement about the bounds --- the physics involved in any specific system very well may imply a dramatically slower heating rate.

Strictly speaking, the above comments only apply within linear response theory.
To go beyond linear response, one could perform a Magnus-like expansion along the lines of Refs.~\cite{Abanin2017Effective,Tran2019Locality}.
However, higher-order terms in the expansion involve higher-body interactions (i.e., terms supported on more than two sites), and so the results of this paper do not immediately apply.
While we expect that our results could be generalized to higher-body interactions, we have not attempted to do so and leave this for future work.

\subsection{Ground state correlations} \label{subsec:ground_state_correlations}

One of the most well-known applications of LR bounds is for the proof that gapped ground states have exponentially decaying correlations.
Considering a 1D nearest-neighbor Hamiltonian $H$ for concreteness (although this result holds much more generally), the statement is that if there is a non-vanishing gap $\Delta E$ between the ground state and first excited state energies, then for any local observables $A_0$ and $B_r$,
\begin{equation} \label{eq:ground_state_clustering_statement}
G(r) \equiv \big< A_0 B_r \big> - \big< A_0 \big> \big< B_r \big> \leq C e^{-\kappa r},
\end{equation}
where expectation values are in the ground state of $H$.
See Refs.~\cite{Hastings2004LiebSchultzMattis,Hastings2006Spectral} for full details of the proof.
Here we only discuss the steps at which LR bounds enter.

The authors of Refs.~\cite{Hastings2004LiebSchultzMattis,Hastings2006Spectral} show that one can bound $G(r)$ by (assuming $\lVert A_0 \rVert = \lVert B_r \rVert = 1$ for simplicity)
\begin{equation} \label{eq:ground_state_clustering_bound_v1}
\big| G(r) \big| \leq \int_{-\infty}^{\infty} \textrm{d}t \frac{1}{|t|} e^{-\frac{t^2 \Delta E^2}{2q}} \Big\lVert \big[ A_0^t, B_r \big] \Big\rVert + 2 e^{-\frac{q}{2}},
\end{equation}
with $q$ to be chosen later.
Much as we described in the previous subsection, assuming a conventional LR bound with velocity $v$, split the integral into one over $|t| < r/2v$ and one over $|t| > r/2v$.
Using the LR bound in the former and the trivial bound $\lVert [A_0^t, B_r] \rVert \leq 2$ in the latter, we find\footnote{\label{note:Hastings_Koma}Note that $[A_0^0, B_r] = 0$ for any $r \neq 0$, and even more, one can show that $[A_0^t, B_r] = O(t)$ at small $t$. Thus the integrand of Eq.~\eqref{eq:ground_state_clustering_bound_v1} does not diverge as $|t| \rightarrow 0$ and the integral is finite. The LR \textit{bound} as presented in this paper, however, does not share the property of vanishing as $|t| \rightarrow 0$, but this can easily be remedied by combining the LR bound with the above observation that $\lVert [A_0^t, B_r] \rVert \leq C|t|$: use the former for $|t| \geq C^{-1}$ and the latter for $|t| < C^{-1}$ (see Refs.~\cite{Hastings2004LiebSchultzMattis,Hastings2006Spectral}).}
\begin{equation} \label{eq:ground_state_clustering_bound_v2}
\begin{aligned}
\big| G(r) \big| &\leq C e^{-\frac{ar}{2}} + C' e^{-\frac{r^2 \Delta E^2}{8v^2 q}} + 2 e^{-\frac{q}{2}}.
\end{aligned}
\end{equation}
Thus if $\Delta E > 0$, setting $q = O(r)$ gives Eq.~\eqref{eq:ground_state_clustering_statement}.

For non-translation-invariant systems, using our modified LR bound, we split the integral over $t$ into additional terms as in the previous subsection (see Eq.~\eqref{eq:modified_distant_response_function_bound}).
Yet we again do not obtain any significant improvement over Eq.~\eqref{eq:ground_state_clustering_bound_v2}, since we still have a term scaling as $\exp{[-ar/2]}$.

In fact, it is unclear whether Eq.~\eqref{eq:ground_state_clustering_bound_v2} itself applies to the systems considered here --- Ref.~\cite{Movassagh2017Generic} proves that ensembles of Hamiltonians are generically \textit{gapless} if the norms of the interactions have continuous distributions extending to zero.
However, the lack of a gap in this case is due to the existence with high probability of nearly-disconnected local regions hosting low-energy excitations~\cite{Movassagh2017Generic}.
Since those excitations are decoupled from the larger system, one does not expect them to give rise to long-range correlations on physical grounds.
To our knowledge, it remains a significant open question whether (and under what conditions) the ground states of such disordered systems have rapidly decaying correlations.

\subsection{Predicting properties of gapped ground states} \label{subsec:predicting_properties}

An interesting recent application in which LR bounds enter is classical machine-learning algorithms for predicting properties of quantum many-body ground states.
The authors of Ref.~\cite{Huang2021Provably} consider a family of time-independent Hamiltonians $H(x)$ parametrized by a continuous (potentially multi-dimensional) variable $x$, with corresponding ground states $\rho(x)$.
They obtain rigorous results on the ability to predict $\textrm{Tr} \rho(x) O$ for a certain $x$ from knowledge of $\{\textrm{Tr} \rho(x_i) O\}_{i=1}^n$ for other parameter values $\{x_i\}_{i=1}^n$, where $O$ is any observable that can be written as a sum of local terms.

A central ingredient is the result that if $H(x)$ has a non-zero spectral gap uniformly in $x$, then one can bound the size of the gradient $\nabla_x \textrm{Tr} \rho(x) O$.
Specializing to 1D chains with $H(x) = \sum_l H_l(x)$ and $O = \sum_i O_i$ (although Ref.~\cite{Huang2021Provably} treats more general systems), the authors show that
\begin{equation} \label{eq:observable_gradient_bound_v1}
\big\lVert \nabla_x \textrm{Tr} \rho(x) O \big\rVert \leq \int_{-\infty}^{\infty} \textrm{d}t W(t) \sum_{il} \Big\lVert \big[ O_i, \hat{u} \cdot \nabla_x H_l^t(x) \big] \Big\rVert,
\end{equation}
where $\hat{u}$ is an arbitrary unit vector in the parameter space, and $W(t)$ is a filter function which decays faster than any polynomial as $|t| \rightarrow \infty$.

Much as before, the conventional LR bound enters by dividing the terms on the right-hand side into two groups, one in which $i$ and $l$ are separated by a distance less than $vt$ and the other in which they are separated by greater than $vt$.
Use the trivial bound on the commutator for the former and the LR bound for the latter.
Since there are $O(|t|)$ terms in the former and the sum over the latter is $O(1)$, Eq.~\eqref{eq:observable_gradient_bound_v1} reduces to
\begin{equation} \label{eq:observable_gradient_bound_v2}
\big\lVert \nabla_x \textrm{Tr} \rho(x) O \big\rVert \leq C \int_{-\infty}^{\infty} \textrm{d}t W(t) |t|.
\end{equation}
The remaining integral is finite (since $W(t)$ decays sufficiently fast), and thus the gradient is bounded.
This property is then used in Ref.~\cite{Huang2021Provably} to establish the efficiency of algorithms capable of predicting $\textrm{Tr} \rho(x) O$.

In a certain sense, our modified LR bound for non-translation-invariant systems provides an improvement, since it allows us to replace the factor of $|t|$ in Eq.~\eqref{eq:observable_gradient_bound_v2} by $|t|^{\alpha}$.
However, the important feature is merely that the resulting integral over $t$ is finite.
Thus while our modified bound does tighten the numerical value of the gradient, it does not seem to give any dramatic changes.
Furthermore, since the analysis of Ref.~\cite{Huang2021Provably} requires that the Hamiltonians $H(x)$ be gapped, the caveats from our discussion of ground state correlations apply here as well.
The more substantive question is whether the conclusions of Ref.~\cite{Huang2021Provably} apply to disordered systems at all.

\section{Acknowledgements} \label{sec:acknowledgements}

It is a pleasure to acknowledge valuable and stimulating discussions with several colleagues: A.\ Chandran, P.\ J.\ D.\ Crowley, A.\ Deshpande, C.\ R.\ Laumann, D.\ M.\ Long, A.\ Polkovnikov, and M.\ C.\ Tran in particular.
We would also like to thank MathOverflow users ``Iosif Pinelis'' and ``mike'' for guiding us through a proof of Eq.~\eqref{eq:threshold_transfer_protocol_runtime}.
We acknowledge funding by the DoE ASCR Quantum Testbed Pathfinder program (award No.~DE-SC0019040), ARO MURI, AFOSR, DoE QSA, NSF QLCI (award No.~OMA-2120757), DoE ASCR Accelerated Research in Quantum Computing program (award No.~DE-SC0020312), NSF PFCQC program, AFOSR MURI, DARPA SAVaNT ADVENT, and the Minta Martin and Simons Foundations.

\bibliography{biblio}

\appendix

\begin{widetext}

\section{Some properties of matrix norms} \label{app:matrix_norms}

In this paper, we use $\lVert \, \cdot \, \rVert$ to denote the operator norm, i.e., the largest eigenvalue of the operator in question (technically the largest singular value, but we shall only apply the norm to Hermitian operators).
It is equivalently the $p \rightarrow \infty$ limit of the Schatten $p$-norm, defined for a generic $M \times M$ matrix $O$ as
\begin{equation} \label{eq:Schatten_norm_definition}
\big\lVert O \big\rVert_p \equiv \left( \frac{1}{M} \textrm{Tr} \big| O \big|^p \right)^{\frac{1}{p}} = \left( \frac{1}{M} \sum_{i=1}^M \big| \lambda_i \big|^p \right)^{\frac{1}{p}}.
\end{equation}
The Frobenius ($p = 2$) norm is often of independent interest (e.g., see Ref.~\cite{Tran2020Hierarchy}).
Let us first note that $\lVert O \rVert_p$ is a non-decreasing function of $p$, and thus the bounds on $\lVert O \rVert$ obtained in this work are automatically bounds on $\lVert O \rVert_p$ for all $p \in (0, \infty)$.
This follows from the fact that for $p < q$, the function $f(x) = x^{q/p}$ is convex on $[0, \infty)$.
Thus
\begin{equation} \label{eq:Schatten_norm_ordering}
\left( \frac{1}{M} \sum_{i=1}^M \big| \lambda_i \big|^p \right)^{\frac{q}{p}} \leq \frac{1}{M} \sum_{i=1}^M \big| \lambda_i \big|^q,
\end{equation}
and exponentiating both sides by $1/q$ gives $\lVert O \rVert_p \leq \lVert O \rVert_q$.

Now restrict ourselves to the operator norm and to tensor product Hilbert spaces, i.e., $M = d^N$.
A useful inequality is Eq.~\eqref{eq:projected_operator_norm_bound} in the main text: for any Hermitian operator $O$ and any subset $\omega$ of the sites, $\lVert \mathcal{P}_{\omega} O \rVert \leq 2 \lVert O \rVert$, where $\mathcal{P}_{\omega}$ projects onto basis strings which act non-trivially somewhere in $\omega$.

To prove this, define $\mathcal{Q}_{\omega} \equiv I - \mathcal{P}_{\omega}$ to be the projector onto strings which \textit{do} act trivially throughout $\omega$.
We have that
\begin{equation} \label{eq:projector_triangle_inequality}
\big\lVert \mathcal{P}_{\omega} O \big\rVert \leq \big\lVert O \big\rVert + \big\lVert \mathcal{Q}_{\omega} O \big\rVert.
\end{equation}
Note that $\mathcal{Q}_{\omega} O$ is a tensor product between $\omega$ and its complement: $\mathcal{Q}_{\omega} O = I_{\omega} \otimes O_{\Omega / \omega}$ for a certain operator $O_{\Omega / \omega}$ acting on $\Omega / \omega$.
Furthermore, $\lVert \mathcal{Q}_{\omega} O \rVert = \lVert O_{\Omega / \omega} \rVert$.
Pick a state $|\psi \rangle$ such that $| \langle \psi | O_{\Omega / \omega} |\psi \rangle| = \lVert O_{\Omega / \omega} \rVert$, and define the normalized density matrix
\begin{equation} \label{eq:projector_trial_state}
\rho \equiv d^{-|\omega|} I_{\omega} \otimes \big| \psi \big> \big< \psi \big|,
\end{equation}
so that $|\textrm{Tr} \rho \mathcal{Q}_{\omega} O| = \lVert \mathcal{Q}_{\omega} O \rVert$.
Also note that $\textrm{Tr} \rho \mathcal{P}_{\omega} O = 0$ (since $\mathcal{P}_{\omega} O$ by definition consists only of basis strings orthogonal to the identity on $\omega$), thus $\lVert \mathcal{Q}_{\omega} O \rVert = |\textrm{Tr} \rho O|$.
Lastly, by the variational principle for density matrices, $|\textrm{Tr} \rho O| \leq \lVert O \rVert$.
Returning to Eq.~\eqref{eq:projector_triangle_inequality}, we have Eq.~\eqref{eq:projected_operator_norm_bound} from the main text.

\section{Proof that projection commutes with decoupled evolution} \label{app:commutation_proof}

Recall the notation of Eq.~\eqref{eq:disconnected_lattice_commutation} in the main text: $\omega$ denotes any subset of sites, and $\lambda$ denotes a subset of links such that $\omega$ and $\Omega / \omega$ are disconnected under $\Lambda / \lambda$.
In words, removing $\lambda$ from the lattice disconnects $\omega$ from everything else.
Eq.~\eqref{eq:disconnected_lattice_commutation}, reproduced here:
\begin{equation} \label{eq:disconnected_lattice_reproduction}
\mathcal{P}_{\omega} \mathcal{U}_{\Lambda / \lambda}(t) = \mathcal{U}_{\Lambda / \lambda}(t) \mathcal{P}_{\omega},
\end{equation}
is the statement that interactions on $\Lambda / \lambda$ alone cannot convert an operator supported outside $\omega$ into one whose support intersects $\omega$, or vice versa.
While obvious on physical grounds, it is worth confirming that this follows from the formal definitions in Sec.~\ref{subsec:definition_basis_strings}.

We need only make the following observations, all of which are clear from the explicit expression for $\mathcal{U}_{\Lambda / \lambda}(t)$ given by Eq.~\eqref{eq:evolution_superoperator_formal_expression}:
\begin{itemize}
\item $\mathcal{U}_{\Lambda / \lambda}(t)$ factors into $\mathcal{U}_{\omega}(t) \otimes \mathcal{U}_{\Omega / \omega}(t)$, where the first factor involves only the interactions within $\omega$ and the second only those outside $\omega$.
\item For any product operator $O \equiv O_{\omega} \otimes O_{\Omega / \omega}$,
\begin{equation} \label{eq:disconnected_evolution_action}
\mathcal{U}_{\Lambda / \lambda}(t) O = \mathcal{U}_{\omega}(t) O_{\omega} \otimes \mathcal{U}_{\Omega / \omega}(t) O_{\Omega / \omega}.
\end{equation}
\item The action of $\mathcal{U}(t)$ takes the identity string to itself and does not take any other basis string onto the identity (this statement is always true for any system).
\end{itemize}
Consider an arbitrary operator $O$, which can always be decomposed as
\begin{equation} \label{eq:generic_operator_basis_decomposition}
O = \sum_{\nu_{\omega}} X_{\omega}^{(\nu_{\omega})} \otimes O_{\Omega / \omega}^{(\nu_{\omega})},
\end{equation}
where $\nu_{\omega}$ denotes a set of basis indices on $\omega$ and $O_{\Omega / \omega}^{(\nu_{\omega})}$ denotes the partial projection of $O$ onto basis string $X_{\omega}^{(\nu_{\omega})}$.
As a result of the above, we have that
\begin{equation} \label{eq:disconnected_lattice_demonstration}
\begin{aligned}
\mathcal{P}_{\omega} \mathcal{U}_{\Lambda / \lambda}(t) O &= \mathcal{P}_{\omega} \sum_{\nu_{\omega}} \mathcal{U}_{\omega}(t) X_{\omega}^{(\nu_{\omega})} \otimes \mathcal{U}_{\Omega / \omega}(t) O_{\Omega / \omega}^{(\nu_{\omega})} \\
&= \sum_{\nu_{\omega} \neq 0} \mathcal{U}_{\omega}(t) X_{\omega}^{(\nu_{\omega})} \otimes \mathcal{U}_{\Omega / \omega}(t) O_{\Omega / \omega}^{(\nu_{\omega})} \\
&= \mathcal{U}_{\Lambda / \lambda}(t) \sum_{\nu_{\omega} \neq 0} X_{\omega}^{(\nu_{\omega})} \otimes O_{\Omega / \omega}^{(\nu_{\omega})} = \mathcal{U}_{\Lambda / \lambda}(t) \mathcal{P}_{\omega} O.
\end{aligned}
\end{equation}

\section{An explicit state transfer protocol} \label{app:transfer_protocol}

The transfer protocol $H(t) \in \mathcal{H}_J$ illustrated in Fig.~\ref{fig:transfer_protocol}, which we use to assess the tightness of our LR bounds in the main text, is given there in terms of SWAP gates.
While perfectly sufficient on its own (see footnote~\ref{note:SWAP} of Sec.~\ref{sec:general_bound}), it is satisfying to see that $H(t)$ can be represented entirely in terms of ``standard'' interactions, even for arbitrary $d$-state degrees of freedom.
This was demonstrated in Ref.~\cite{GarciaEscartin2013Swap}, and we summarize their construction here for the sake of completeness.

The circuit diagram for a single SWAP gate is shown in Fig.~\ref{fig:SWAP_gate}.
The central element is the add-invert gate $\widetilde{X}_{ij}$ (subscripts indicating the two sites involved), defined as
\begin{equation} \label{eq:add_invert_gate_definition}
\widetilde{X}_{ij} \big| q_i, q_j \big> \equiv \big| q_i, -q_j - q_i \big>,
\end{equation}
where $q_i, q_j \in \{0, 1, \cdots, d-1\}$ label single-site basis states.
All arithmetic here is to be interpreted mod $d$.
Then $\textrm{SWAP}_{ij} = \widetilde{X}_{ij} \widetilde{X}_{ji} \widetilde{X}_{ij}$:
\begin{equation} \label{eq:swap_gate_demonstration}
\big| q_i, q_j \big> \stackrel{\widetilde{X}_{ij}}{\longrightarrow} \big| q_i, -q_j - q_i \big> \stackrel{\widetilde{X}_{ji}}{\longrightarrow} \big| q_j, -q_j - q_i \big> \stackrel{\widetilde{X}_{ij}}{\longrightarrow} \big| q_j, q_i \big>.
\end{equation}
The construction of the add-invert gate involves the single-site Fourier Transform gate,
\begin{equation} \label{eq:fourier_transform_gate_definition}
\textrm{FT}_j \big| q_j \big> \equiv \frac{1}{\sqrt{d}} \sum_{p_j=0}^{d-1} e^{2\pi i q_j p_j/d} \big| p_j \big>,
\end{equation}
and the controlled-Z gate,
\begin{equation} \label{eq:controlled_Z_gate_definition}
\textrm{CZ}_{ij} \big| q_i, q_j \big> \equiv e^{2\pi i q_i q_j/d} \big| q_i, q_j \big>.
\end{equation}
One can directly confirm that $\widetilde{X}_{ij} = \textrm{FT}_j \textrm{CZ}_{ij} \textrm{FT}_j$~\cite{GarciaEscartin2013Swap}.

\begin{figure}[t]
\centering
\includegraphics[width=0.7\textwidth]{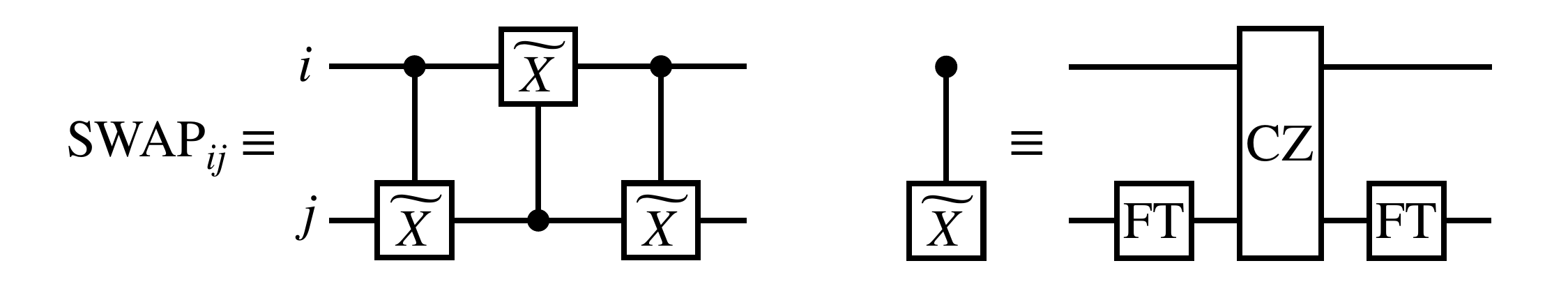}
\caption{Implementation of a SWAP gate between sites $i$ and $j$. The $\widetilde{X}_{ij}$ gate is given in Eq.~\eqref{eq:add_invert_gate_definition}, the $\textrm{FT}_j$ gate is given in Eq.~\eqref{eq:fourier_transform_gate_definition}, and the $\textrm{CZ}_{ij}$ gate is given in Eq.~\eqref{eq:controlled_Z_gate_definition}.}
\label{fig:SWAP_gate}
\end{figure}

We need to express this procedure in terms of a Hamiltonian $H(t) \in \mathcal{H}_J$.
The only interaction required is for implementation of the controlled-Z gate.
Defining the local operator $Z_j \equiv \textrm{diag}(0, 1, \cdots, d-1)$, it is simply a $Z_i Z_j$ interaction.
In order to respect the constraint on link $l$ that $\lVert H_l \rVert \leq J_l$ (note that $\lVert Z_j \rVert = d-1$), we set
\begin{equation} \label{eq:controlled_z_gate_interaction_term}
H_l = -\frac{J_l}{(d-1)^2} Z_{l-1} Z_l.
\end{equation}
The controlled-Z gate on link $l$ then amounts to applying interaction $H_l$ for time $2\pi (d-1)^2/dJ_l$ (and three such gates are needed per SWAP gate).
As for the FT gates, since arbitrary local terms are allowed in the family $\mathcal{H}_J$ (see footnote~\ref{note:local_terms} of Sec.~\ref{sec:summary}), each amounts to applying a local field for some fixed time.
The total runtime required to effect state transfer from site 0 to $r$ is therefore (not including the $O(r)$ runtime from all FT gates)
\begin{equation} \label{eq:transfer_protocol_precise_runtime}
T_r = \frac{6\pi (d-1)^2}{d} \sum_{l=1}^r \frac{1}{J_l}.
\end{equation}

\section{Some probabilistic tools} \label{app:probabilistic_tools}

Here we summarize some of the mathematical tools from probability theory needed in Sec.~\ref{sec:disordered_bound} of the main text.
Far more detail can be found in textbooks on the subject, such as Refs.~\cite{Rosenthal2006,Durrett2019} (including derivations of the following, which we shall not present here).

To begin, keep in mind that the fundamental objects in probability theory are \textit{subsets} of the set of all possible outcomes, called ``events'', together with a function $\textrm{Pr}$ that maps such subsets to the interval $[0, 1]$.
Oftentimes not all subsets can be included in the domain of $\textrm{Pr}$, and $\textrm{Pr}$ must obey certain natural properties, but we will not dwell on these here.
The important thing to note is simply that we can perform all of the usual set-theoretic operations on events, such as the union or intersection, and many of the basic tools in probability theory involve relating the values of $\textrm{Pr}$ with respect to those operations.

Our first tool is ``countable subadditivity'' (also known as the ``union bound''): for any countable (potentially infinite) collection of events $\{E_r\}$,
\begin{equation} \label{eq:countable_subadditivity}
\textrm{Pr} \left[ \bigcup_r E_r \right] \leq \sum_r \textrm{Pr} \big[ E_r \big].
\end{equation}
We will primarily use Eq.~\eqref{eq:countable_subadditivity} after establishing that $\textrm{Pr}[E_r] = 0$ for all $r$ --- it then follows that $\textrm{Pr}[\cup_r E_r] = 0$.
In words, if each event $E_r$ separately has probability zero, then the probability that any of them occur (even if there are a countably infinite number) is also zero.

The next tool is the ``continuity'' of probabilities.
Suppose that $\{A_r\}$ is an ``increasing'' set of events, in that $A_1 \subseteq A_2 \subseteq \cdots$, and similarly that $\{B_r\}$ is a ``decreasing'' set of events, in that $B_1 \supseteq B_2 \supseteq \cdots$.
We then have that
\begin{equation} \label{eq:continuity_probabilities}
\textrm{Pr} \left[ \bigcup_r A_r \right] = \lim_{r \rightarrow \infty} \textrm{Pr} \big[ A_r \big], \qquad \textrm{Pr} \left[ \bigcap_r B_r \right] = \lim_{r \rightarrow \infty} \textrm{Pr} \big[ B_r \big].
\end{equation}
For our purposes, we will have events either of the form $A_r =$ ``the event that $E_{r'}$ occurs for all $r' > r$'' or $B_r =$ ``the event that $E_{r'}$ occurs for some $r' > r$''.
In terms of set-theoretic operations, these are given by
\begin{equation} \label{eq:increasing_decreasing_events}
A_r \equiv \bigcap_{r' > r} E_{r'}, \qquad B_r \equiv \bigcup_{r' > r} E_{r'}.
\end{equation}
Note that $\{A_r\}$ is an increasing set of events and $\{B_r\}$ is a decreasing set.
We will specifically want to determine the probability that some $A_r$ occurs and the probability that all $B_r$ occur.
The former event is denoted ``$E_{r'}$ a.a.''\ with a.a.\ abbreviating ``almost always'', and the latter event is denoted ``$E_{r'}$ i.o.''\ with i.o.\ abbreviating ``infinitely often''.
These events are given by
\begin{equation} \label{eq:aa_io_set_definitions}
E_{r'} \textrm{ a.a.} \equiv \bigcup_r \bigcap_{r' > r} E_{r'}, \qquad E_{r'} \textrm{ i.o.} \equiv \bigcap_r \bigcup_{r' > r} E_{r'}.
\end{equation}
Note that $E_{r'}$ a.a.\ and $E_{r'}$ i.o.\ can be described in words as ``there exists an $r$ past which all $E_{r'}$ occur'' and ``past every $r$ there is some $E_{r'}$ that occurs'', i.e., ``$E_{r'}$ occurs almost always'' and ``$E_{r'}$ occurs infinitely often'' respectively, in exactly the same manner as we have used in discussing LR bounds.
Continuity --- Eq.~\eqref{eq:continuity_probabilities} --- allows us to express the probabilities of these events as
\begin{equation} \label{eq:continuity_probabilities_example}
\textrm{Pr} \big[ E_{r'} \textrm{ a.a.} \big] = \lim_{r \rightarrow \infty} \textrm{Pr} \left[ \bigcap_{r' > r} E_{r'} \right], \qquad \textrm{Pr} \big[ E_{r'} \textrm{ i.o.} \big] = \lim_{r \rightarrow \infty} \textrm{Pr} \left[ \bigcup_{r' > r} E_{r'} \right].
\end{equation}

We lastly need the ``first Borel-Cantelli lemma'', which says that if a collection of events $\{E_r\}$ has probabilities such that $\sum_r \textrm{Pr}[E_r]$ is finite, then the probability is 1 that only a finite number of the events occur:
\begin{equation} \label{eq:Borel_Cantelli}
\sum_r \textrm{Pr} \big[ E_r \big] < \infty \quad \longrightarrow \quad \textrm{Pr} \big[ E_r \textrm{ i.o.} \big] = 0.
\end{equation}
In particular, suppose we have a sequence of random variables $\{X_r\}$ and want to prove that it converges to zero with probability 1.
The definition of $\{X_r\}$ \textit{not} converging to zero is that there exist some $\epsilon > 0$ such that for every $R$, $|X_r| > \epsilon$ for some $r > R$.
In other words, $|X_r| > \epsilon$ i.o.\ for all $\epsilon > 0$.
It suffices to consider only rational $\epsilon > 0$, and thus by countable subadditivity, we only need to prove that $\textrm{Pr}[|X_r| > \epsilon \textrm{ i.o.}] = 0$ for each individual $\epsilon$.
By the Borel-Cantelli lemma, it further suffices (but need not be necessary) to prove that $\sum_r \textrm{Pr}[|X_r| > \epsilon]$ is finite.
If each individual term $\textrm{Pr}[|X_r| > \epsilon]$ can be calculated or at least bounded directly, this is a very useful line of attack.

\end{widetext}

\end{document}